\documentclass[12pt]{JHEP}
\usepackage{latexsym,amssymb,psfrag}
\usepackage{graphicx}
\usepackage{pstricks}
\usepackage{amsmath}
\usepackage{subfigure}

\title{Neutralino Dark Matter beyond CMSSM Universality}

\author{Vincent Bertin $^{1}$, Emmanuel Nezri $^{1}$ $^{2}$,
  Jean Orloff $^{2}$ \\
  $^{1}$Centre de Physique des Particules de Marseille\\
  IN2P3-CNRS, Universit\'e de la M\'editerran\'ee, 
  F-13288 Marseille Cedex 09\\
  $^{2}$Laboratoire de Physique Corpusculaire de Clermont-Ferrand\\
  IN2P3-CNRS, Universit\'e Blaise Pascal, F-63177 Aubiere Cedex}

\abstract{ We study the effect of departures from SUSY GUT universality on
  the neutralino relic density, and both its direct and indirect
  detection, especially by neutrino telescopes. We find that the most
  interesting models are those with a value of $M_3|_{GUT}$ lower than the
  universal case.}

\preprint{PCCF-RI-0216\\
CPPM-P-2002-02}


\begin{document}

\section{Introduction - CMSSM summary}

In a supersymmetric framework with $R-$parity conservation, the lightest
supersymmetric particle (LSP) is stable. In the Minimal Supersymmetric
Standard Model (MSSM), it is often the lightest neutralino ($\equiv$ {\it
  the} neutralino $\chi$) which is a neutral Majorana particle. It then
offers an interesting candidate to account for cold dark matter (CDM) in
the present Universe ($\Omega_{CDM}\sim 0.3$). The relic population of
these neutralinos which survive after spatial separation freezes their
self-annihilation, could be detected by the energy they transfer to nuclei
in direct detection experiments. Another possibility is the indirect
detection of fluxes coming from the decays of neutralino annihilation
products ($\chi\chi\rightarrow X\bar{X}\rightarrow \nu,\gamma,e^+\ ...$)
(for a review on neutralino dark matter and different detection
possibilities see \cite{Jungman:1996df}).  Since these fluxes go like the
squared neutralino density, they require some concentration to restart
their annihilation. The gravitational concentration achieved in galactic
halos is much too low for neutrino indirect detection. However,
neutralinos can also accumulate in the gravitational well of massive
astrophysical bodies, and the larger concentration achieved in the centre
of the Sun can give rise to higher muon neutrino fluxes detectable by
neutrino telescopes such as Antares \cite{antaresweb} or IceCube
\cite{Icecube}.

In a previous paper \cite{Bertin:2002ky} we studied the potential detection
of neutralino dark matter by neutrino telescopes and direct detection
experiments in CMSSM models (also known as mSugra). Those models inspired from
minimal supergravity and gravity-mediated SUSY breaking assume a
unification of the soft parameters of the MSSM at high energy $M_{GUT}\sim
2\times 10^{16}$ GeV reducing the 106 ``SUSY'' parameters of the MSSM down
to 5 : universal masses for scalars $(m_0)$ and gauginos $(m_{1/2})$,
universal trilinear $(A_0)$ and bilinear $(B_0)$ couplings, and a Higgs
``mass'' parameter $(\mu_0)$. Using renormalisation group equations (RGE)
and requiring radiative electroweak supersymmetry breaking, the usual input
parameters are $m_0$, $m_{1/2}$, $A_0$, $\tan{\beta}$ (the ratio of the 2
Higgs doublet vacuum expected values at low energy) and ${\rm
sign}(\mu)$. The neutralinos are the mass eigenstates coming from the
mixing of neutral gauge and Higgs boson superpartners. In the MSSM the
neutralino mass matrix in the
($\tilde{B},\tilde{W}^3,\tilde{H}^0_1,\tilde{H}^0_2$) basis is :

\begin{equation}
M_N=\left( \begin{array}{cccc}
M_1 & 0 & -m_Z\cos \beta \sin \theta_W^{} & m_Z\sin \beta \sin \theta_W^{}
\\
0 & M_2 & m_Z\cos \beta \cos \theta_W^{} & -m_Z\sin \beta \cos \theta_W^{}
\\
-m_Z\cos \beta \sin \theta_W^{} & m_Z\cos \beta \cos \theta_W^{} & 0 & -\mu
\\
m_Z\sin \beta \sin \theta_W^{} & -m_Z\sin \beta \cos \theta_W^{} & -\mu & 0
\end{array} \right)
\label{eq:matchi}
\end{equation}
and can be diagonalised by a single mixing matrix $z$ : $M_{\rm diag} = z M_N z^{-1}$.
The (lightest) neutralino is then given by the linear combination
\begin{equation}
\chi = z_{11}\tilde{B}     +z_{12} \tilde{W}^3
          +z_{13}\tilde{H}^0_1 +z_{14} \tilde{H}^0_2.
\end{equation}

\begin{figure}[t!]
\begin{center}
\begin{tabular}{cc}
 \includegraphics[width=0.5\textwidth]{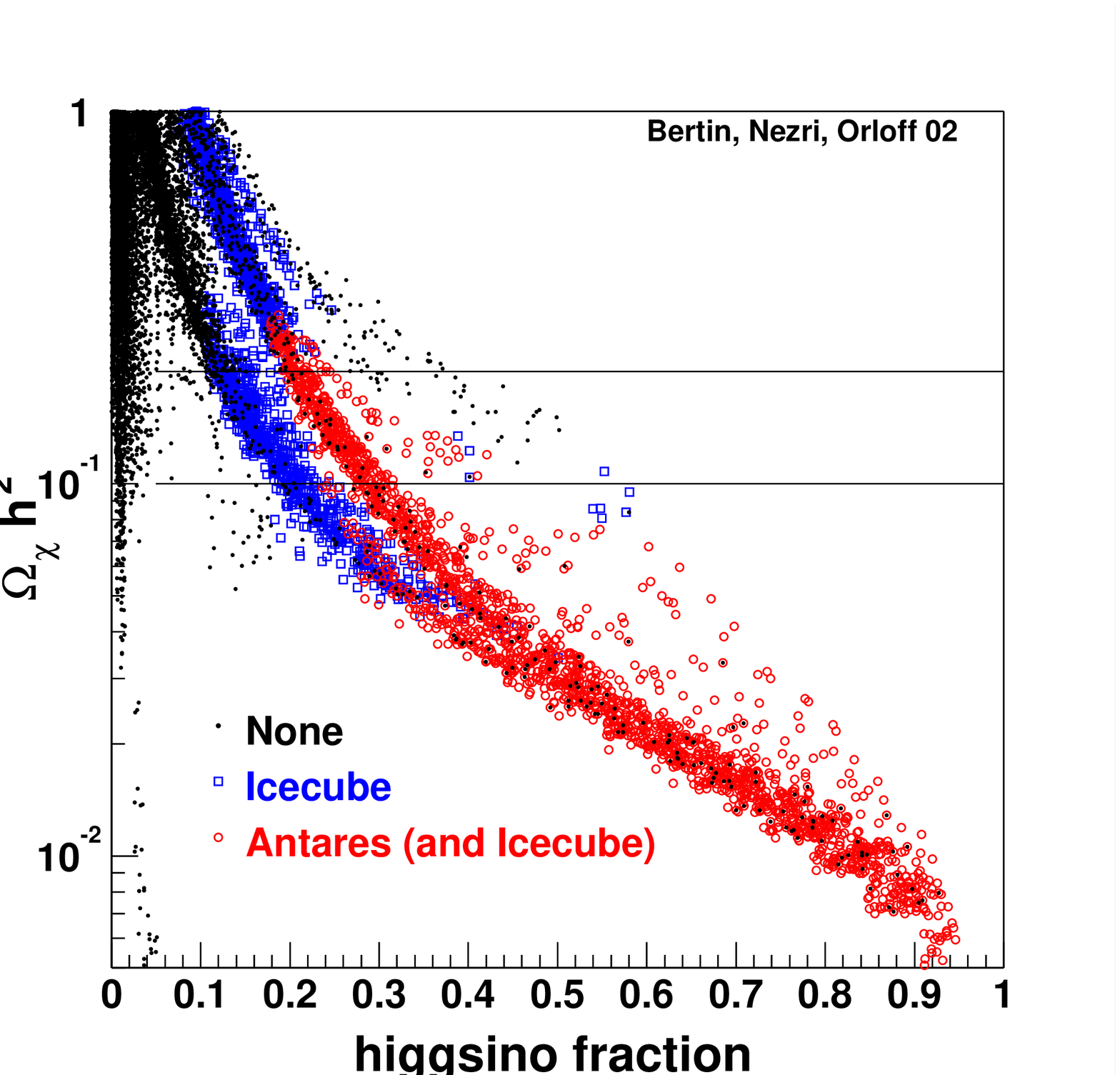} &
\includegraphics[width=0.5\textwidth]{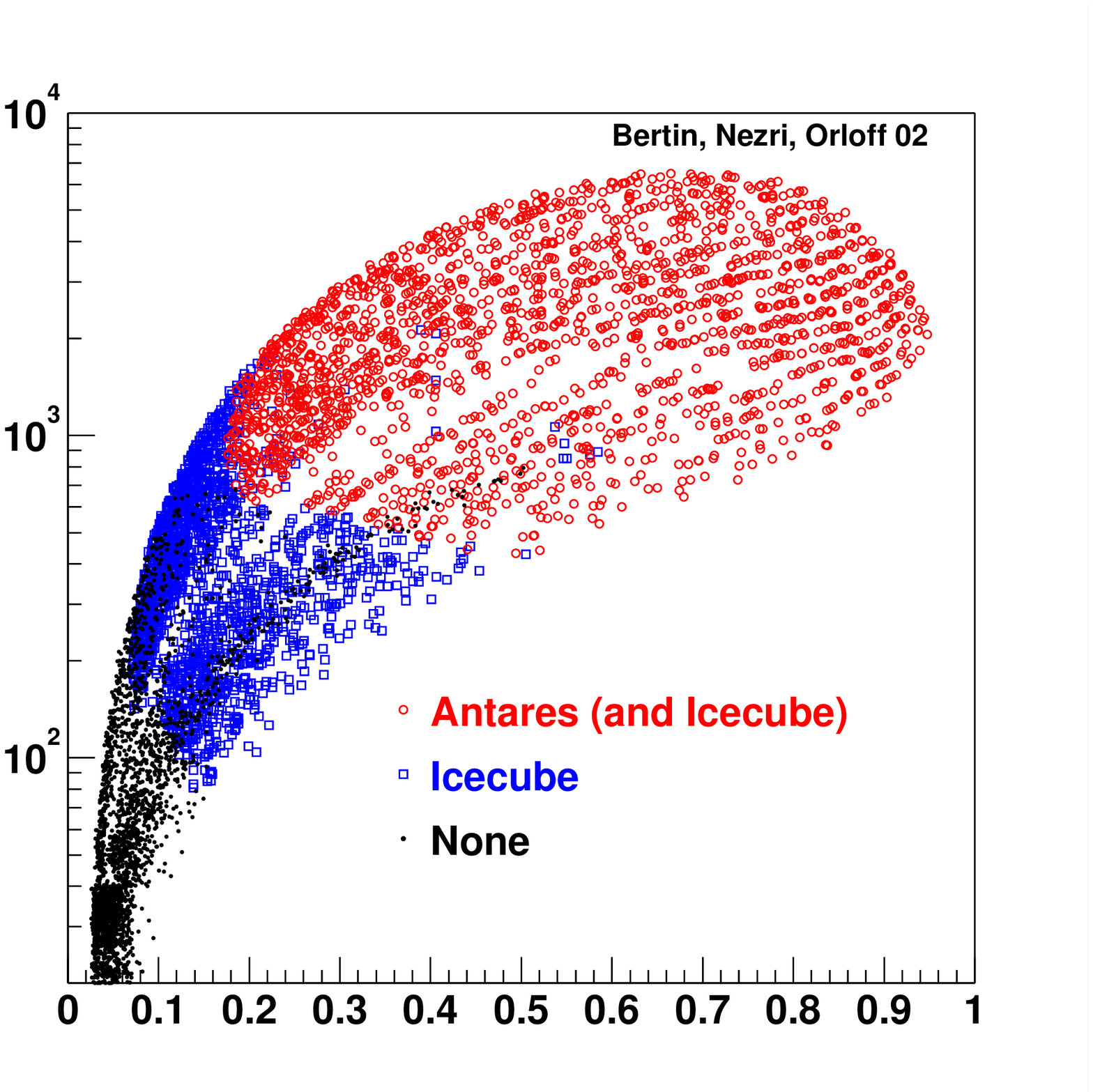}\\ a) & b)
\end{tabular}
\caption{\small a) Neutralino relic density and b) muon flux coming
from the Sun as functions of the $\chi$ higgsino fraction for a wide
scan of CMSSM models (see text for details). The band in plot a)
indicates the current preferred range for $\Omega_{CDM}$.}
\label{resumemSugra}
\end{center}
\end{figure}

In CMSSM models, the neutralino exhibits two different natures: it is fully
bino-like $(z_{11}\approx 1)$ for low $m_0$ values because of RGE evolution
of parameters in eq.(\ref{eq:matchi}), and can acquire a non-negligible
higgsino component for $m_0\gtrsim 1000$ GeV along the boundary where
radiative electroweak symmetry breaking cannot be achieved because of the
focus point behaviour \cite{Feng:1999zg}. In a previous work
\cite{Bertin:2002ky}, we used {\tt Suspect}\footnote{Suspect has been
  recently updated. This improvement does not affect our results.}
\cite{Suspect,Djouadi:2002ze} for RGE, potential minimisation and SUSY
spectrum calculations, and {\tt DarkSusy}\cite{Darksusy} to estimate relic
density and detection rates. We carefully analysed the dominant
annihilation channels which count both in relic density calculation and
neutrino spectra. In regions where the neutralino relic density satisfy
current cosmological constraints, we found that:
\begin{itemize}
\item the low $m_0$ and $m_{1/2}$ region is strongly constrained by
  experimental limits on SUSY contributions to the $b\rightarrow s+\gamma$
  branching ratio, on the lightest Higgs mass and on SUSY contributions to
  $(g-2)_{\mu}$ : $a^{SUSY}_{\mu}$,
\item $\chi\tilde{t}$ and $\chi\tilde{\tau}$ coannihilation regions are
  beyond reach of detection,
\item the  pseudo-scalar $A$ pole region can only be potentially detected 
with very big future projects for direct detection ($\sim$ 1
ton size),
\item the large $m_0$ ``focus point'' region where the neutralino has a
  significant (and crucial, see figure \ref{resumemSugra}b ) higgsino
  fraction ($f_H$) is interesting for direct and indirect detection
  experiments. Neutrino/muon fluxes coming from the Sun are large only in
  this region because of the $\chi\chi\xrightarrow{\chi^+_i,\chi_i}
  W^+W^-,\ ZZ$ and $\chi\chi\xrightarrow{Z}t\bar{t}$ channels which give
  rise to more energetic neutrino spectra and to muons with higher energy.
\end{itemize}

\begin{figure}[t!]
\begin{tabular}{cc}
 \includegraphics[width=0.5\textwidth]{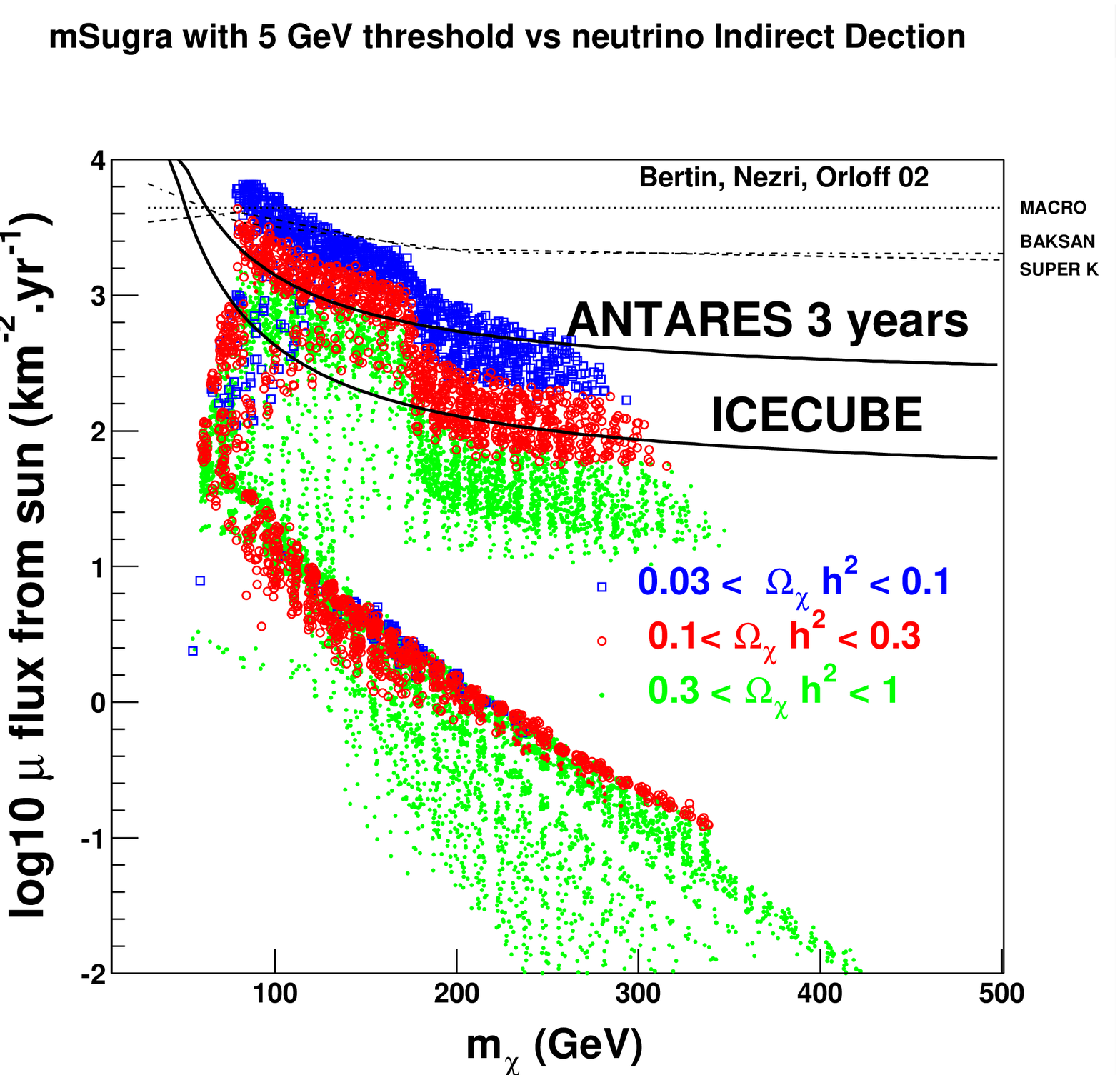}&
\includegraphics[width=0.5\textwidth]{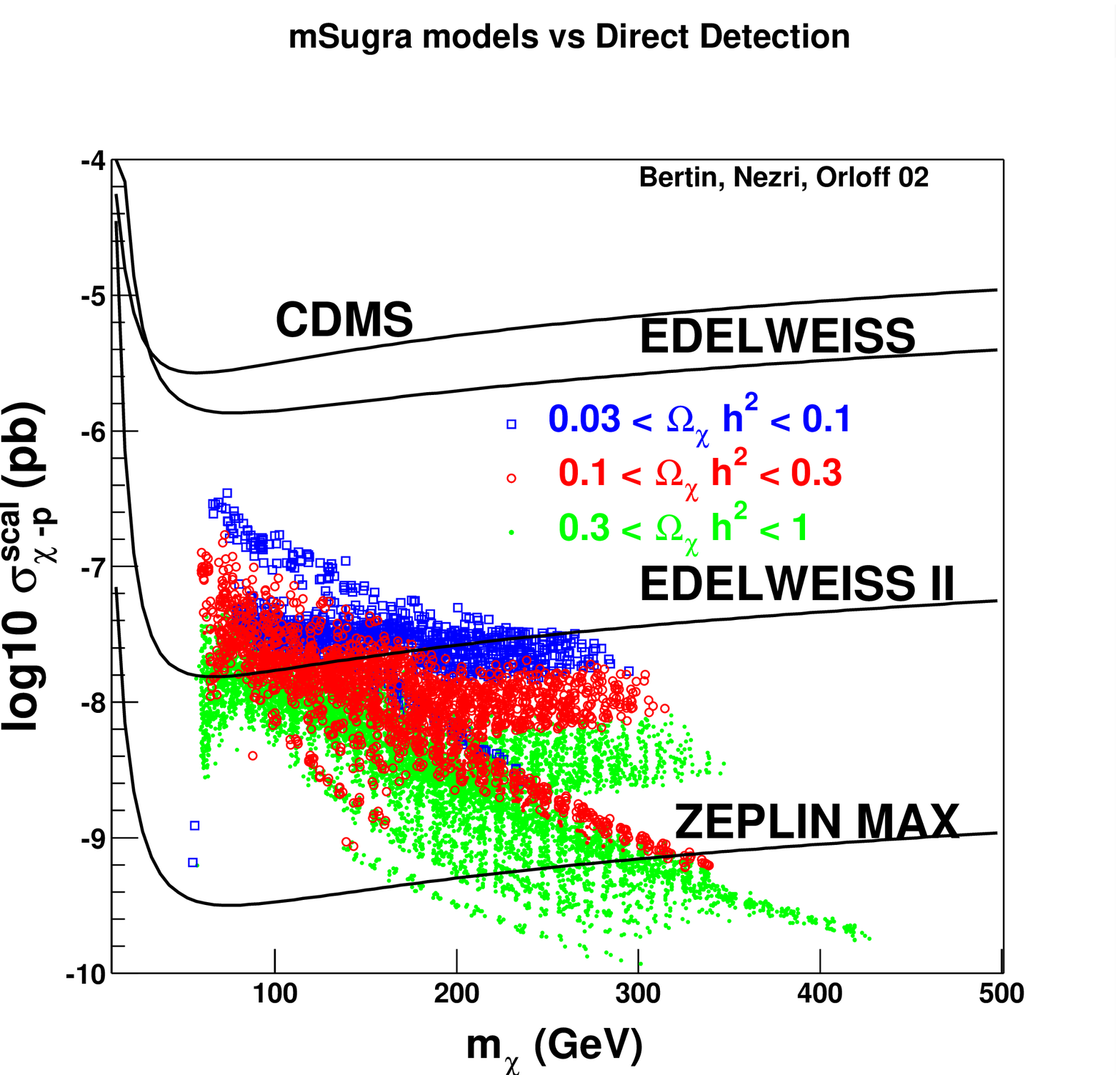}\\
a) & b)
\end{tabular}
\caption{\small 
  a) Neutrino indirect detection experimental (Macro \cite{Macro}, Baksan
  \cite{Suvorova:1999my}, Super-Kamiokande \cite{SuperK}, Antares
  \cite{AntarLee}, IceCube \cite{Ice3Edsjo}) sensitivities on muon fluxes
  with a 5 GeV threshold coming from $\chi$ annihilations in the Sun and b)
  direct detection experimental (CDMS \cite{Abusaidi:2000wg}, Edelweiss
  I\cite{Benoit:2002hf} and II\cite{EdelweissII}, Zeplin \cite{Zeplin})
  sensitivities on $\sigma^{scal}_{\chi-p}$.  Both are function of the
  neutralino mass for a wide scan of CMSSM models (see text for details).}
\label{mSugrasensitiv}
\end{figure}

This is summarised on figures \ref{mSugrasensitiv} and \ref{resumemSugra}
for the following sample of CMSSM models, compared to sensitivities of
current and future direct and neutrino indirect detection experiments :
\begin{itemize}
\item $0<m_0<3000$ GeV ; $40<m_{1/2}<800$ GeV ; $A_0=0$ GeV ;
  $\tan{\beta}=10,\ 50$ ; $\mu>0$,
\item $0<m_0<3000$ GeV ; $40<m_{1/2}<1000$ GeV ; $A_0=-800,\ -400,\ 0,\ 
  400,\ 800$ GeV ; $\tan{\beta}=20,\ 35$ ; $\mu>0$,
\item $0<m_0<3000$ GeV ; $40<m_{1/2}<1000$ GeV ; $A_0=0$ GeV ;
  $\tan{\beta}=45$ ; $\mu>0$.
\end{itemize}

\noindent
The constraints we apply on the models are :
\begin{itemize}
\item limits on SUSY particle masses from accelerators searches (see e.g.
  \cite{SUSY02susy} to
  update these values): \\
  $m_{\chi_1^+}>104$ GeV; $m_{\tilde{f}}>100$ GeV for
  $\tilde{f}=\tilde{t}_1,\tilde{b}_1,\tilde{l}^{\pm},\tilde{\nu}$,
  $m_{\tilde{g}}>300$ GeV; $m_{\tilde{q}_{1,2}}>260$ GeV for
  $\tilde{q}=\tilde{u},\tilde{d},\tilde{s},\tilde{c}$,
\item limits on Higgs mass \cite{SUSY02Higgs} : we require $m_h>112$ GeV
  instead of $m_h>114$ GeV (depending on $\tan(\beta-\alpha)$) because of
  the too low value for Higgs mass in Suspect 2.005 ($\sim 3$ GeV) as
  pointed out in \cite{Allanach:2002pz},
\item limits on $b\rightarrow s\gamma$ branching ratio : we require
  $BR(b\to s+\gamma)=1 \to 4 \times 10^{-4}$ as calculated in DarkSusy,
  corresponding to $BR(b\to s+\gamma)\simeq2.3 \to 5.3 \times 10^{-4}$ (the
  Standard Model value calculated in DarkSusy is $2.4\times 10^{-4}$
  instead of the NLO $3.6\times 10^{-4}$ \cite{Ciuchini:1998xe}) when
  experimental results are \cite{PDG} $BR(b\to s+\gamma)=3.37 \pm 0.37 \pm
  0.34\pm 0.24\pm 0.38 \times 10^{-4}$.
\item limits on SUSY contribution to muon anomalous moment $(g-2)_{\mu}$
  \cite{Knecht:2001qf} :
  $-6\times10^{-10}<a^{SUSY}_{\mu}<58\times10^{-10}$.
\end{itemize}

The Renormalisation Group Equations (RGE) lead SUSY models to a generic
hierarchy of particle spectrum in which scalars are heavier than light
neutralinos and charginos.  As summarised above, the most
interesting CMSS Models for the detection of a relic neutralino are those for
which its higgsino fraction is non negligible and has a dominant effect
(figure \ref{resumemSugra}). These are indeed the only models leading to
large neutralino annihilation cross section $\sigma_{\chi-\chi}^A$
(important for its relic density and for indirect detection) {\it and} to
large neutralino--proton scalar and spin dependent elastic cross sections
(important for direct and indirect detection). In this paper, using the
same tools as previously \cite{Suspect,Darksusy}, we shall relax some
universality hypotheses and examine whether more favourable models for
relic density and detections (both direct and neutrino telescopes) can be
found, and establish that models detectable by neutrino telescopes are more
generic than the ``focus point'' region of a typical $(m_0,m_{1/2})$ CMSSM
plane, offering a less constrained framework for detection. We will then
conclude by a generic low energy parameterisation of typical models coming
from RGE evolutions, favourable for neutrino indirect detection.

\section{Non-universal SUSY GUT models}
\label{nonuniv}

In which direction can one relax the universality hypothesis in order to
induce the physics we want, namely to get large $\sigma^A_{\chi-\chi}$,
$\sigma^{scal}_{\chi-q}$ and $\sigma^{spin}_{\chi-q}$ and thus a non
negligible higgsino fraction for the neutralino ?

We can start with some arguments on non-universality inspired by the
(1-loop) renormalisation group equations \cite{Kazakov:1999pe}.  It has
been shown \cite{Kazakov:1998uj} that from the RGE solutions of the
couplings in the unbroken symmetry phase, one can obtain the soft terms of
the broken phase by an expansion over Grassman variables.  This has been
done for the MSSM couplings \cite{Kazakov:1999pe} in order to get the RGE
solutions for the soft terms. In the case of an analysis with three
independent Yukawa couplings $h_t$, $h_b$ and $h_{\tau}$, it is found that
due to the infra-red quasi fixed point (IRQFP) \cite{Hill:1985tg} behaviour
of the Yukawa couplings, the low energy values of the third generation
scalar soft masses and of the Higgs masses have a weak (scale independent)
relation with their initial values at $M_{GUT}$ and depend mainly on the
high energy $SU(3)$ gaugino soft mass $M_3|_{GUT}$ (in a scale dependent
way).  The first and second generation soft masses have an analogous
behaviour because of their negligible Yukawa couplings. This can be written
as :
\begin{equation}
  (M^{scal}_{soft}|_{low})^2= 
  (M^{scal}_{soft}|_{GUT})^2+c_3f_3+c_2f_2+c_1f_1+corrections\ 
\label{eq:nonuniv}
\end{equation} 
with
\begin{equation}
f_i=\frac{(M_i^{GUT})^2}{b_i}\left( 1-\frac{1}{(1+b_i\alpha_0
t)^2}\right) ,
\end{equation}
where $M_i$ are the soft gaugino mass terms and $\alpha_0$ is the
universal gaugino coupling at $M_{GUT}$, and where $c_3$ is strongly
dominant for squarks and Higgses and less dominant for sleptons which
have vanishing $SU(3)$ charges.

In addition, a gaugino non-universality given by
$(M_2/M_1)_{GUT}<1$ can lead to a large wino component for the
neutralino and to an important modification of its couplings with
respect to the universal CMSSM case and thus to a very
different phenomenology.

We will now explore various non-universal scenarios in order to
study their possible benefits on detection.

\section{Non-universality of scalar soft masses and of trilinear
couplings at $M_{GUT}$}

{\bf Sfermions :}\\
As for $A_0$ in the CMSSM \cite{Bertin:2002ky}, modifications of the soft
parameters in the sfermions mass matrices can give rise to light third
generation sfermions. This could modify the neutralino relic density value
through coannihilation processes. However, in detection processes, the
neutralino interacts with nuclei and thus mainly with $u$ and $d$ valence
quarks. Due to their small Yukawa couplings, RGE evolutions of the first
and second generation squark masses only depend on gaugino soft masses.
This implies that their masses can not be lowered by changing scalar soft
terms to enhance $\sigma^{scal}_{\chi-q}$ and $\sigma^{spin}_{\chi-q}$
through the process $\chi q\xrightarrow{\tilde{q}}\chi q$. Non-universality
in the sfermion mass soft terms can thus only lead to $\chi\tilde{\tau}$
\cite{Ellis:1999mm}, $\chi\tilde{t}$
\cite{Boehm:1999bj,Djouadi:2001yk,Ellis:2001nx} coannihilation effects,
giving rise to models for which the neutralino is cosmologically a good
dark matter candidate without modifying its detectability.

The soft terms for the third generation sfermions being present in soft
Higgs masses RGE's ($X_t$, $X_b$ and $X_{\tau}$) can also influence the
radiative electroweak symmetry breaking through the running of $m^2_{H_1}$
and $m^2_{H_2}$. This effect is similar to the one discussed in the next
section. One could also add lighter third generation squarks but this is
not helpful for detection.

\begin{figure}[t!]
\begin{center}
\begin{tabular}{cc}
\multicolumn{2}{c}{{ $m_{H_2}=m_0\ ;\ m_{H_1}=0.5m_0\ ;\ A_0=0\ ;\ \tan{\beta}=45\ ;\ \mu>0$}}\\
\multicolumn{2}{c}{\includegraphics[width=\textwidth]{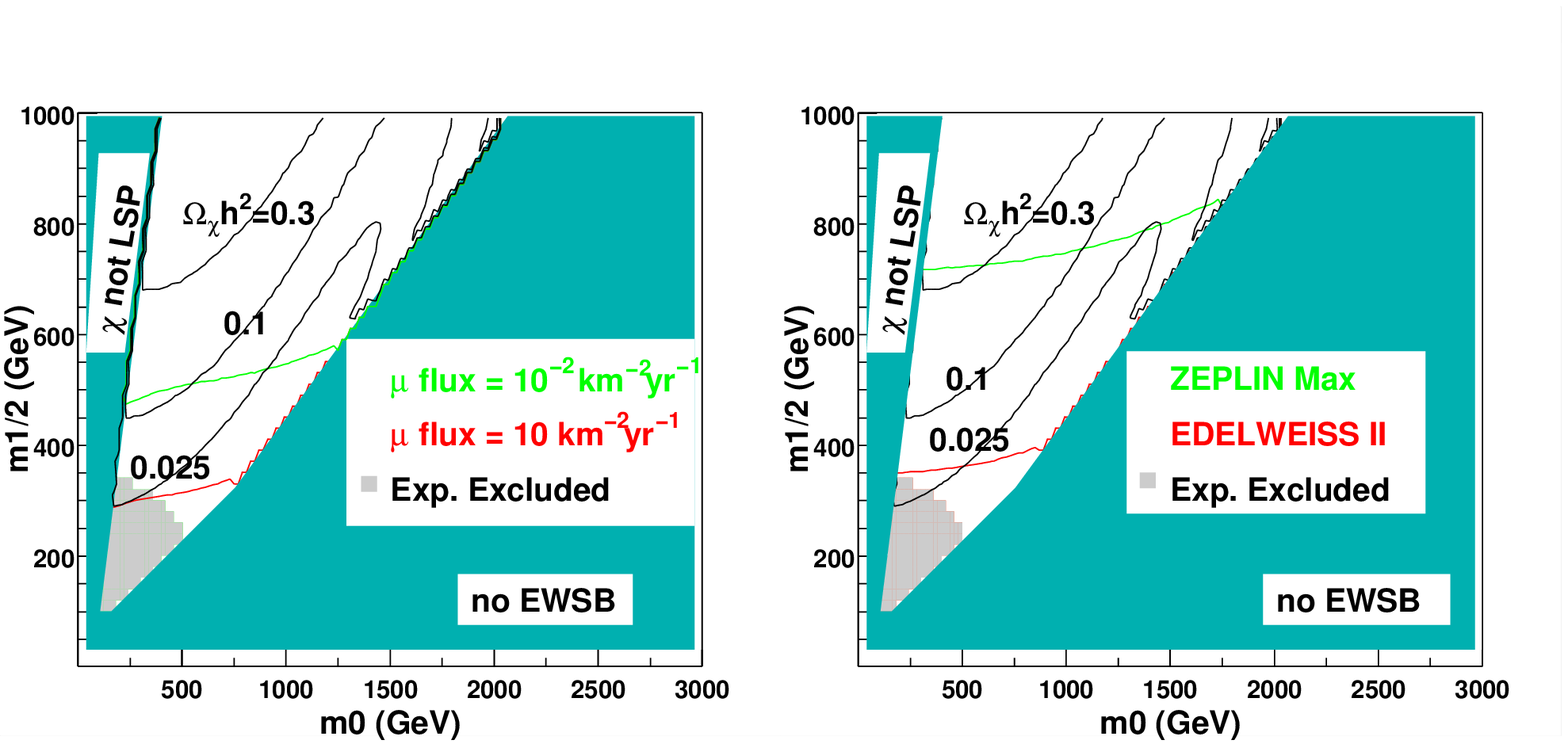}}\\
\hspace{0.25\textwidth}a) & \hspace{0.15\textwidth}b)
\end{tabular}
\caption{\small Neutralino detection potential a) on muon fluxes coming
  from the Sun for neutrino telescopes and b) for direct detection
  experiments in the $(m_0,m_{1/2})$ plane for models with
  non-universal Higgs soft masses $m^2_{H_2}=m_0$ and $m^2_{H_1}=0.5 m_0$
  and with $A_0=0$, $\tan{\beta}=45$, $\mu>0$. Also shown are constant relic
  density lines $\Omega_{\chi}h^2=1,\ 0.3,\ 0.1,\ 0.025$. The small grey
  area shows the models excluded by current experimental constraints.}
\label{nonunivHiggs}
\end{center}
\end{figure}

\noindent {\bf Higgses :}\\ One can relax the universality relation for the
Higgs soft masses at the Grand Unification scale
$m_{H_1}|_{GUT}=m_{H_2}|_{GUT}=m_0$. $m_{H_1}$ and $m_{H_2}$ being now free
parameters, the potential minimisation and thus the value of $\mu$ is less
constrained. The parameterisation usually taken is
\cite{Berezinsky:1996ga,Ellis:2002iu,Barger:2001ur,Ellis:2002wv} :
\begin{equation}
  m_{H_i}|_{GUT}=(1+\delta_i)m_0\ ;\ {\rm for}\ i=1,2.
\end{equation} 
Due to the departure of the Higgs soft masses with respect
to the universal case, $\mu$ values are modified and the parameter space of
a typical $(m_0,m_{1/2})$ plane satisfying the radiative electroweak
symmetry breaking is reduced with respect to the CMSSM. In the remaining
parameter space, models with smaller masses for the heavier Higgs bosons
$m_A$ and $m_H$ are more easily obtained than for the (universal) CMSSM,
leading to wider zones with a good neutralino relic density and accessible
direct detection yields. Concerning the neutrino indirect detection
potential, the muon fluxes due to neutralino annihilation in the Sun remain
however small for such models. This is in contradiction with the recent results
presented in ref.\cite{Barger:2001ur}, in which the modification of the
Higgs soft mass relations generate more easily models with large muon
fluxes coming from the Sun in the parameter space allowed by radiative
electroweak symmetry breaking. This difference can certainly be attributed
to the use in \cite{Barger:2001ur} of IsaSusy (version $<7.64$) for RGE's
and SUSY spectrum calculation (instead of Suspect in this work) for which
the predicted value of $\mu$ is usually too small for high $m_0$ values
\cite{Allanach:2002pz}, and hence the higgsino fraction of the neutralino
is too large.  An example of our results is shown on figure
\ref{nonunivHiggs} for $\delta_1=-0.5$ and $\delta_2=0$.

\section{Non-universality of gaugino soft masses at $M_{GUT}$}

\subsection{$SU(5)$ representations}
In CMSS Models, a universal mass is given to all gaugino fields at the
high energy scale $M_{GUT}$. The unification gauge group should either be
$SU(5)$ or a larger group naturally breaking into an $SU(5)$ subgroup at a
scale around $M_{GUT}$. In the class of models for which the SUSY breaking
is induced by an $F$-term, the gaugino masses are generated by a chiral
superfield with an auxiliary component $F_{\Phi}$ acquiring a vev such that
\cite{Amundson:1996nw,Nath:1997qm,Corsetti:2000yq}
\begin{equation}
{\cal M}_{gauginos}\sim\frac{<F_{\Phi}>_{ab}}{M_{Planck}}\lambda^a\lambda^b
\end{equation}
where $\lambda^a$ and $\lambda^b$ are the gaugino fields $\tilde{B}$,
$\tilde{W}$ and $\tilde{g}$.  Since all gaugino fields belong to the
adjoint representation, $F_{\Phi}$ belongs to an irreducible representation
in the symmetrised product of two adjoints of $SU(5)$ (={\bf 24}) or to a
linear combination of these representations :
\begin{equation}
({\bf 24}\times{\bf 24})_{sym}={\bf 1}\oplus{\bf 24}\oplus{\bf 75}\oplus{\bf 200}
\end{equation} 
where only the singlet component ({\bf 1}) leads to universal masses 
$M_1|_{GUT}=M_2|_{GUT}=M_3|_{GUT}=m_{1/2}$. 
The embedding coefficients of the Standard Model gauge groups in
$SU(5)$ give the relations between the gaugino masses at $M_{GUT}$
(see table~\ref{SU5relations}), which we now discuss for the pure
non-singlet irreducible representations.

\begin{table}
\begin{center}
\begin{tabular}{|c|ccc|ccc|}
\hline
\ & \multicolumn{3}{c|} {$M_{GUT}$} & \multicolumn{3}{c|}{$m_Z$} \cr
$F_{\Phi}$
& $M_3$ & $M_2$ & $M_1$
& $M_3$ & $M_2$ & $M_1$ \cr
\hline
${\bf 1}$ & $1$ &$\;\; 1$ &$\;\;1$ & $\sim \;6$ & $\sim \;\;2$ &
$\sim \;\;1$ \cr
${\bf 24}$ & $2$ &$-3$ & $-1$ & $\sim 12$ & $\sim -6$ &
$\sim -1$ \cr
${\bf 75}$ & $1$ & $\;\;3$ &$-5$ & $\sim \;6$ & $\sim \;\;6$ &
$\sim -5$ \cr
${\bf 200}$ & $1$ & $\;\; 2$ & $\;10$ & $\sim \;6$ & $\sim \;\;4$ &
$\sim \;10$ \cr
\hline
\end{tabular}
\caption{\small Relative values of the gaugino masses at $M_{GUT}$ and $m_Z$
  scales in the four possible irreducible representations for $F_{\Phi}$.}
\label{SU5relations}
\end{center}
\end{table}

\noindent
{\bf Case of the 24 :}\\
For low $m_0$ values, since
$M^{24}_1/M^{24}_2<M^{CMSSM}_1/M^{CMSSM}_2\sim1/2$ at low energy, the
neutralino is strongly bino-like. As for the CMSSM, this kind of models will
not lead to large indirect detection muon fluxes coming from the Sun. In
addition, the lower limits on the neutralino mass obtained by LEP
experiments are only valid for $M_1/M_2\geq1/2$ at low energy, so models
with a very light neutralino are not excluded. But the main grey bubble
part of the exclusion domain comes from the lower limits on the SUSY
contribution to the muon $g-2$ parameter $a^{SUSY}_{\mu}$ which are very
strong for these models.

\begin{figure}[t!]
\begin{center}
\begin{tabular}{cc}
\multicolumn{2}{c}{{ ${\bf 24}\ :\  A_0=0\ ;\ \tan{\beta}=45\ ;\ \mu>0$}}\\
\multicolumn{2}{c}{\includegraphics[width=\textwidth
  ]{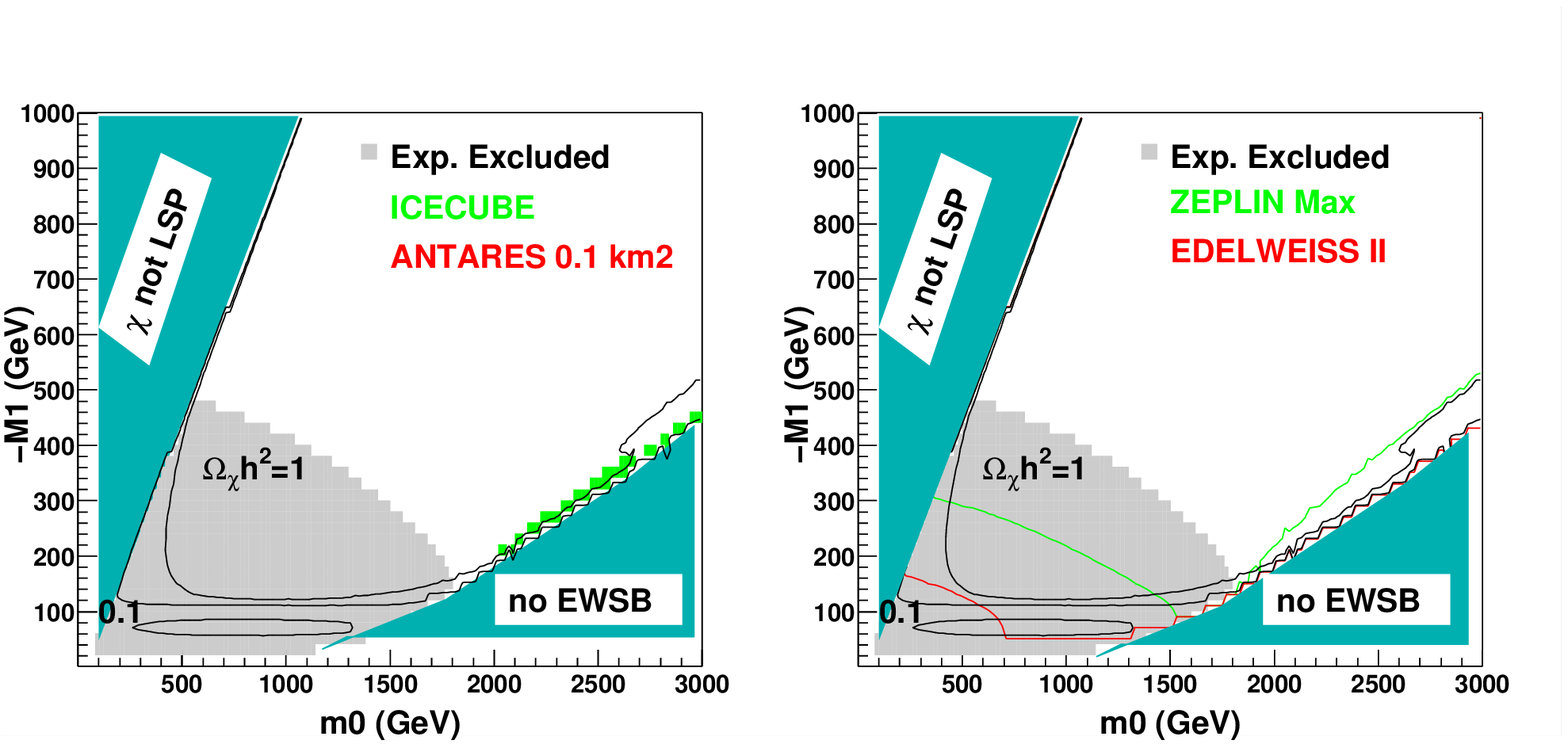}}\\
\hspace{0.25\textwidth}a) & \hspace{0.15\textwidth}b)
\end{tabular}
\caption{\small Neutralino detection potential a) for neutrino telescopes
  detecting muon fluxes coming from the Sun and b) for direct detection
  experiments in the $(m_0,-M_1)$ plane of the {\bf 24} representation
  for $A_0=0$, $\tan{\beta}=45$, $\mu>0$. No model with a good relic
  density is accessible to Antares (0.1 km$^2$).}
\label{nonuniv24}
\end{center}
\end{figure}

For large $m_0$ values, a small higgsino fraction in the neutralino can be
obtained in the focus point region, but less easily than in the CMSSM
universal case because the larger $|M_2|$ value leads to larger $m^2_{H_2}$
values. So the direct detection yields and indirect detection muon fluxes
for these models are less important than for the universal case and the
neutralino relic density is too large. This is illustrated in
figure~\ref{nonuniv24}, where no model is accessible to the Antares
sensitivity, and where the very thin IceCube detection region always leads
to a cosmic over-weight. In addition, the Edelweiss II and Zeplin detection
regions are strongly reduced with respect to the universal case for large
$m_0$ values, due to the small higgsino fraction of the neutralino, and
also due to the relative sign between $M_1$, $M_2$ and $\mu$ (same effect
as $\mu<0$ for the CMSSM with the ``hole'' in the spin independent elastic
cross section due to the suppression between up and down quarks
contributions \cite{Ellis:2000ds}).

This is in agreement with the results of ref.\cite{Belanger:2000tg} in
which the neutralino has been chosen even more bino-like, $M_2/M_1\sim10$,
leading to a relic over-density except for light sleptons.

\noindent
{\bf Case of the 75 :}\\
At low energy $|M_1|\sim|M_2|\sim|M_3|$, the neutralino is thus equally
wino and bino, and the two lightest neutralinos and the lightest chargino
are almost degenerate $M_{\chi} \sim M_{\chi_2^0} \sim M_{\chi^+_1}$. These
models open up the very efficient $\chi\chi\xrightarrow{\chi^+_1,\ 
  \chi^0_2} W^+W^-,\ ZZ$ annihilation channels, as well as very strong
$\chi\chi^0_2$ and $\chi\chi^+_1$ coannihilations.  This leads to a strong
reduction of the relic density. In addition, the gluino can sometimes be
the LSP.

\noindent
{\bf Case of the 200 :}\\ 
Here $|M_2|\sim2/5|M_1|$ at low energy
giving a strongly wino-like neutralino.  
$\chi\chi\xrightarrow{\chi^+_1,\ \chi^0_2} W^+W^-,\ ZZ$ annihilations and
 $\chi\chi^+_1$ coannihilation totally suppress the neutralino
population. 

\subsection{Free relations in gaugino mass parameters : effect of $M_3|_{GUT}$}

As explained above, the MSSM non-universality parameters giving the
strongest impact on the detection of dark matter neutralinos are
$M_2|_{GUT}$ and mostly $M_3|_{GUT}$. We will now study the departure from
universality of these two parameters and their benefits on the neutralino
relic density and the detection yields. These non-universal values will
then be translated in the above $SU(5)$ representation decomposition.

\begin{figure}[ht!]
\begin{center}
\begin{tabular}{ccc}
\multicolumn{3}{c}{{ $m_0=1500\ ;\ m_{1/2}=600\ ;\  A_0=0\ ;\ 
    \tan{\beta}=45\ ;\ \mu>0$}}\\
\hspace{2.8cm}a) & \hspace{4.5cm} b) & \hspace{1.5cm} c)\\
\multicolumn{3}{c}{\includegraphics[width=\textwidth]
  {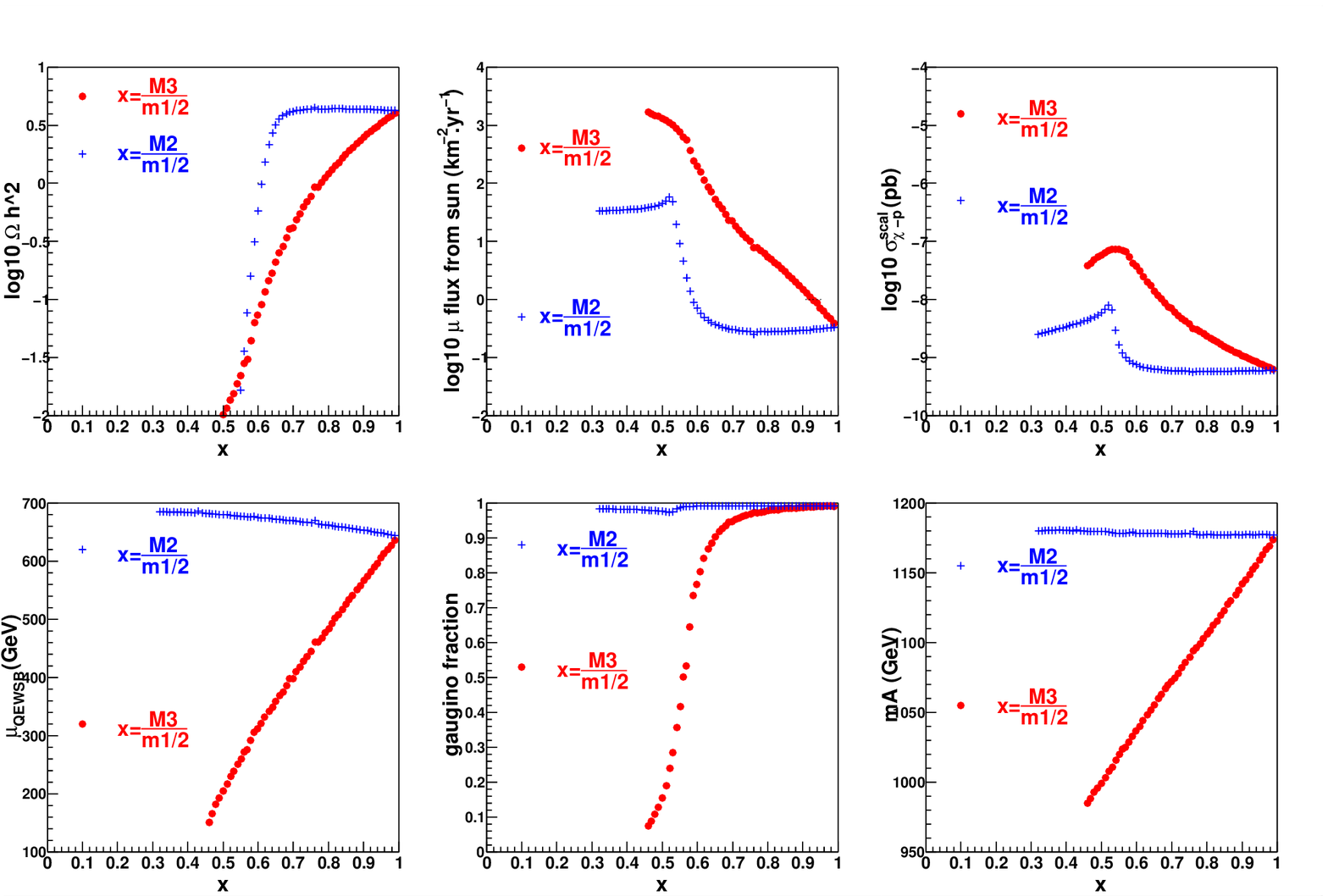}}\\
\hspace{2.8cm}d) & \hspace{4.5cm}e) & \hspace{1.5cm} f)
\end{tabular}
\caption{\small Evolution of a) the neutralino relic density, b) the
  muon flux coming from the Sun, c) the spin independent
  neutralino-proton cross section (direct detection), d) the $\mu$
  parameter, e) the gaugino fraction  and f) the pseudo-scalar mass
  $m_A$ (f) as functions of the $\frac{M_2}{m_{1/2}}$ and
  $\frac{M_3}{m_{1/2}}$ ratios for the CMSS Model with $m_0=1500$ {\rm
    GeV}, $m_{1/2}=600$ {\rm GeV}, $A_0=0$ {\rm GeV}, $\tan{\beta}=45$,
  $\mu>0$.}
\label{M3M2effect}
\end{center}
\end{figure}

\begin{figure}[ht!]
\begin{center}
\begin{tabular}{ccc}
\multicolumn{3}{c}{{ $m_0=1500\ ;\ m_{1/2}=600\ ;\  A_0=0\ ;\ \tan{\beta}=10\ ;\ \mu>0$}}\\
\hspace{2.8cm}a) & \hspace{4.5cm} b) & \hspace{1.5cm} c)\\
\multicolumn{3}{c}{\includegraphics[width=\textwidth]
  {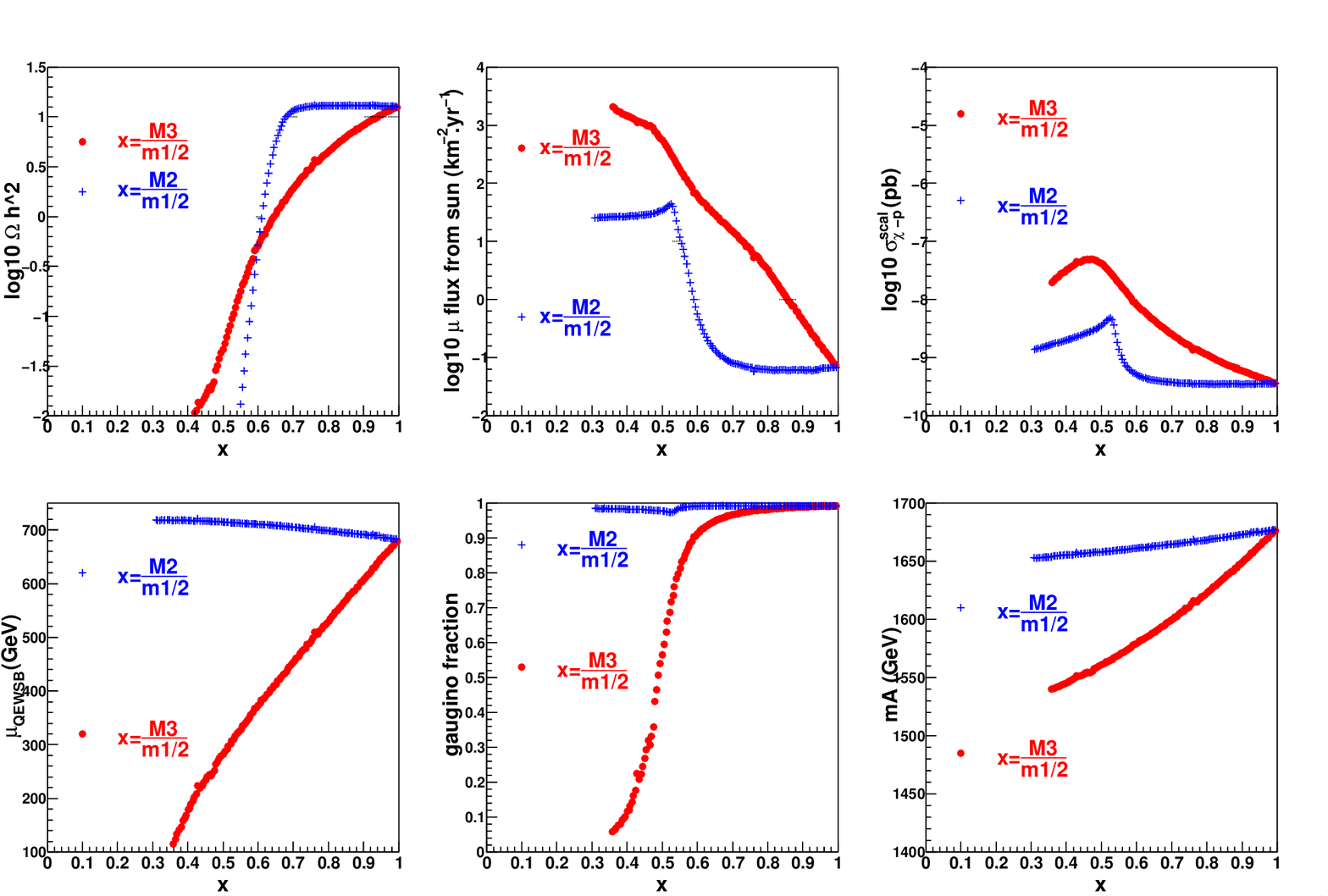}}\\
\hspace{2.8cm}d) & \hspace{4.5cm}e) & \hspace{1.5cm} f)
\end{tabular}
\caption{\small Same as figure \ref{M3M2effect} for $\tan{\beta}=10$
  instead of 45.}

\label{M3M2effect3}
\end{center}
\end{figure}

In the following, the departure from universality effects are
quantified by the ratios at $M_{GUT}$ $x=M_2/m_{1/2}$ or $x=M_3/m_{1/2}$ (with
$m_{1/2}=M_1=M_3$ or $m_{1/2}=M_1=M_2$ respectively) which will be
lowered starting from the CMSSM case ($x=1$).

\begin{figure}[ht!]
\begin{center}
\begin{tabular}{ccc}
\multicolumn{3}{c}{{ $m_0=3000\ ;\ m_{1/2}=2000\ ;\  A_0=0\ ;
    \ \tan{\beta}=45\ ;\ \mu>0$}}\\
\hspace{2.8cm}a) & \hspace{4.5cm} b) & \hspace{1.5cm} c)\\
\multicolumn{3}{c}{\includegraphics[width=\textwidth]
  {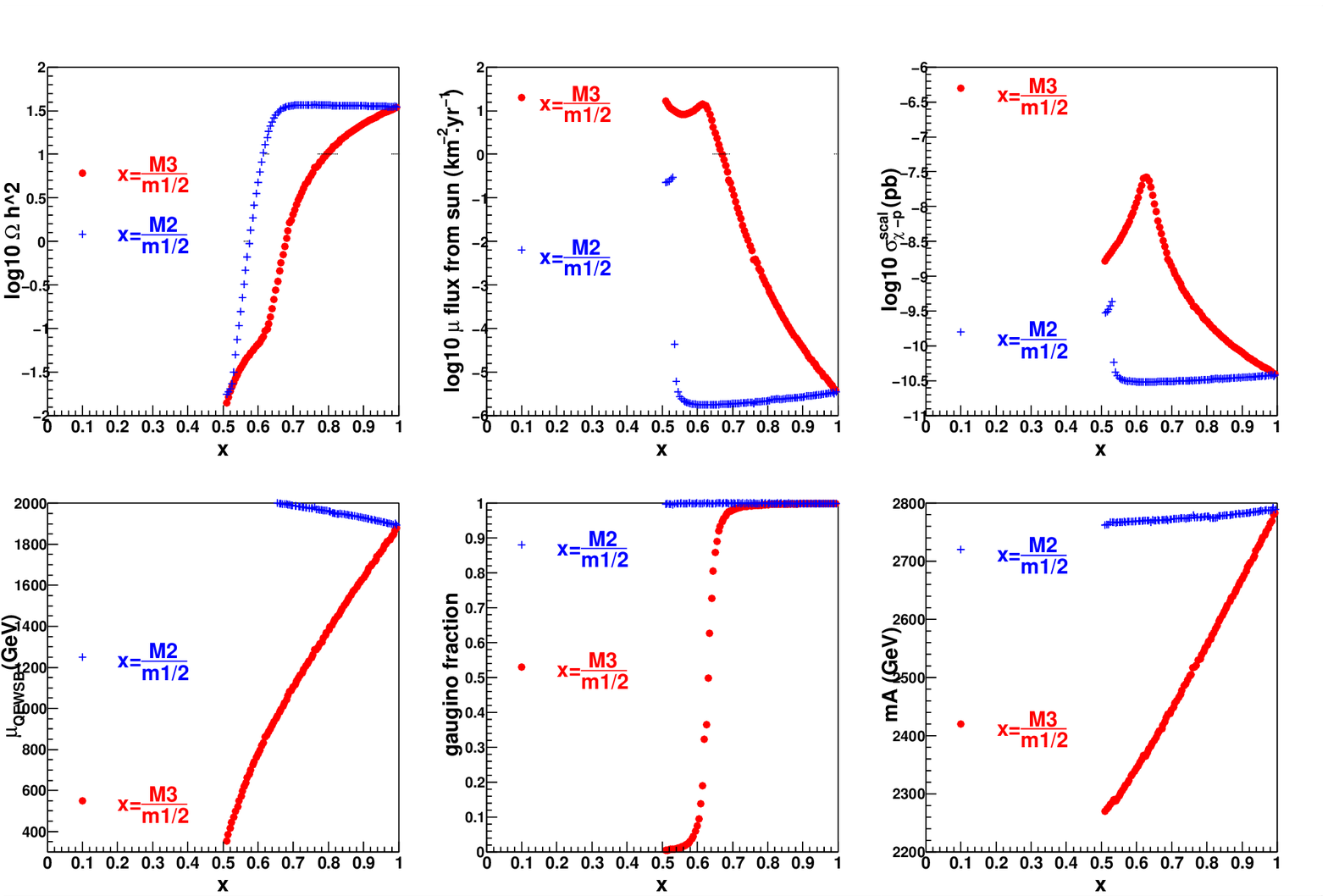}}\\
\hspace{2.8cm}d) & \hspace{4.5cm}e) & \hspace{1.5cm} f)
\end{tabular}
\caption{\small Same as figure \ref{M3M2effect} for the extreme CMSS
  Model with $m_0=3000$ {\rm GeV}, $m_{1/2}=2000$ {\rm GeV}, $A_0=0$ {\rm
    GeV}, $\tan{\beta}=45$, $\mu>0$.}
\label{M3M2effect2}
\end{center}
\end{figure}

\noindent
{\bf The $M_2|_{GUT}$ parameter:}\\
The effect of the $M_2$ parameter is essentially a modification of the
neutralino composition. When the wino component of the neutralino
increases, the $\chi\chi\xrightarrow{\chi^+_1,\ \chi^0_2} W^+W^-,\ ZZ$
processes become more effective and enhance the annihilation cross section
$\sigma^A_{\chi-\chi}$ \cite{Birkedal-Hansen:2001is}. In addition, the
strong $\chi\chi^0_2$ and $\chi\chi^+_1$ coannihilations become active and
the neutralino relic density strongly decreases. This wino component can
increase detection rates by an order of magnitude at most. Indeed, the
neutralino-quark coupling, which enters in direct detection
($\sigma^{scal}_{\chi-q}$, $H$ exchange) and in the capture for indirect
detection ($\sigma^{spin}_{\chi-q}$, $\tilde{q}$ exchange), is
$\tan\theta_W$-suppressed for pure bino w.r.t. pure wino.  The neutralino
annihilations into the hard $W^+W^-$ spectrum also give rise to more
energetic muons.  These enhancements of the (in)direct detection yields can
reach several order of magnitude with respect to the CMSSM case depending
on the values of the other SUSY parameters as can be seen in figures
\ref{M3M2effect}b and \ref{M3M2effect}c, \ref{M3M2effect3}b and
\ref{M3M2effect3}c, and \ref{M3M2effect2}b and \ref{M3M2effect2}c.

However, the relevant value of $M_2$ is very critical, so its benefits are
only operative in a very narrow range. Given an $M_2/m_{1/2}\sim 0.6-0.7$
ratio (equivalent to $M_1|_{low}\sim M_2|_{low}$), the neutralino detection
yields are enhanced but the relic density drops down to much too small
values (see figures \ref{M3M2effect}a, \ref{M3M2effect3}a and
\ref{M3M2effect2}a ). In conclusion, the handling of this $M_2$ parameter
in order to get the desired neutralino dark matter phenomenology can only
be done by ``fine-tuning''. One way around this wino neutralino
extermination suggested in \cite{Moroi:1999zb} is to have the wino
neutralino population derived from AMSB models ($M_1|_{low}\simeq
3M_2|_{low}$) regenerated at low temperature by moduli decays which could
give a good relic abundance.

\noindent
{\bf The $M_3|_{GUT}$ parameter:}\\
The impact of variations in the $M_3$ parameter is much more interesting.
It is indeed one of the key parameters of the MSSM through the RGE's. Its
influence goes well beyond the neutralino sector. Indeed, following the RGE
\cite{Kazakov:1999pe} (see equation \ref{eq:nonuniv}), a decrease of
$M_3|_{GUT}$ leads to a decrease of $m^2_{H_u}$ and of $\mu$ through the
radiative electroweak symmetry breaking mechanism, thus enhancing the
neutralino higgsino fraction, and leads also to a decrease of
$m_{\tilde{q}}$ and $m_A$. These effects are illustrated on the figures
\ref{M3M2effect}d,\ref{M3M2effect}e,\ref{M3M2effect}f,
\ref{M3M2effect3}d,\ref{M3M2effect3}e,\ref{M3M2effect3}f and
\ref{M3M2effect2}d,\ref{M3M2effect2}e,\ref{M3M2effect2}f. The neutralino
relic density then gradually decreases with $x=M_3/m_{1/2}$ (see figures
\ref{M3M2effect}a, \ref{M3M2effect3}a and \ref{M3M2effect2}a ) due to the
increase of the CMSSM dominant annihilation cross section channel (mainly
$\chi\chi\xrightarrow{A}b\bar{b}$, $\chi\chi\xrightarrow{Z}t\bar{t}$,
$\chi\chi\xrightarrow{\chi^+_i} W^+W^-$ and $\chi\chi\xrightarrow{\chi_i}
ZZ$ according to the CMSSM starting parameters \cite{Bertin:2002ky}). The
annihilation channels which directly depend on the neutralino higgsino
fraction finally dominate when $x$ is further lowered, because of the
decreasing of $\mu$.

\begin{figure}[ht!]
\begin{center}
\begin{tabular}{ccc}
\includegraphics[width=0.33\textwidth]{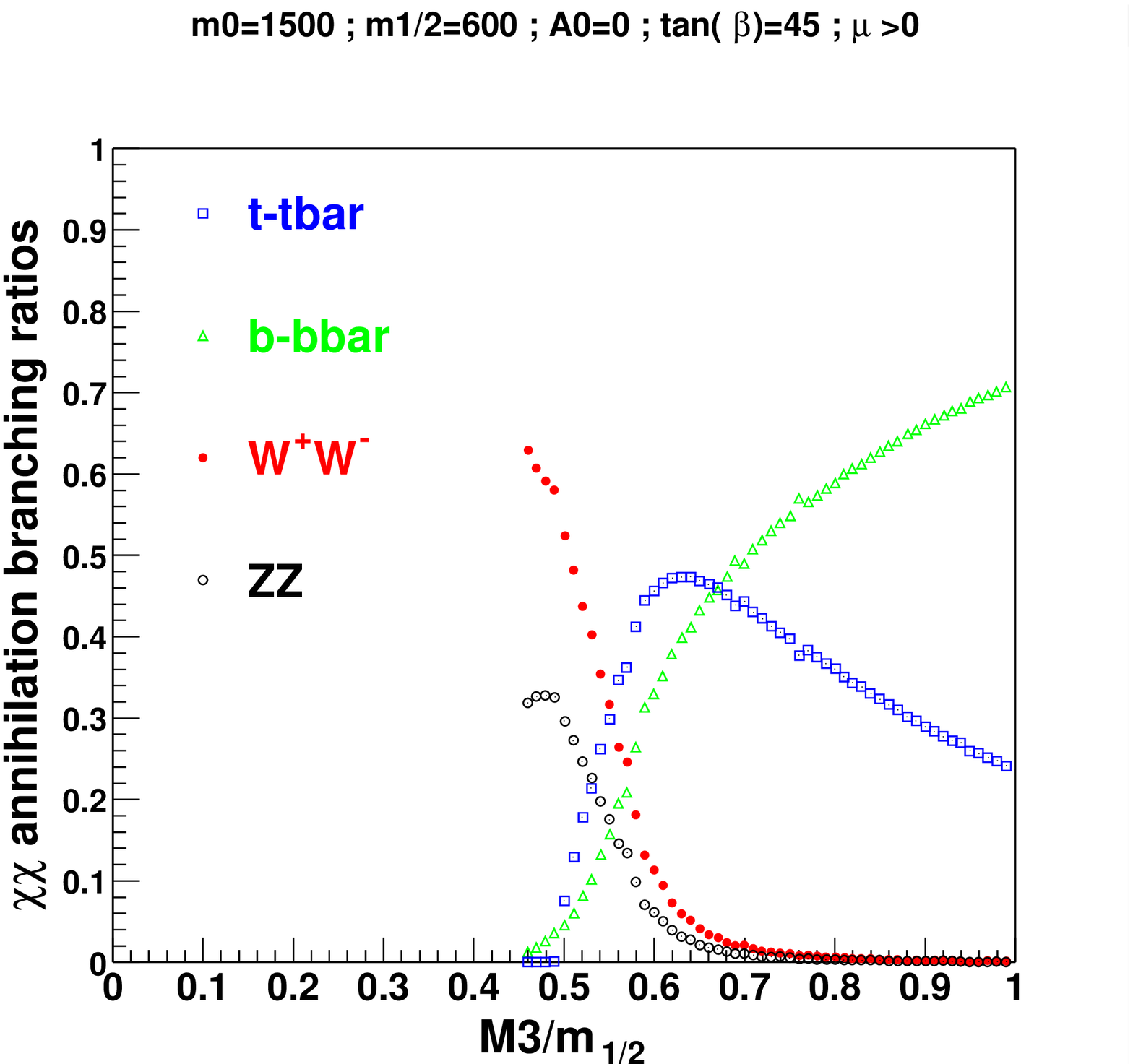} &
\includegraphics[width=0.33\textwidth]{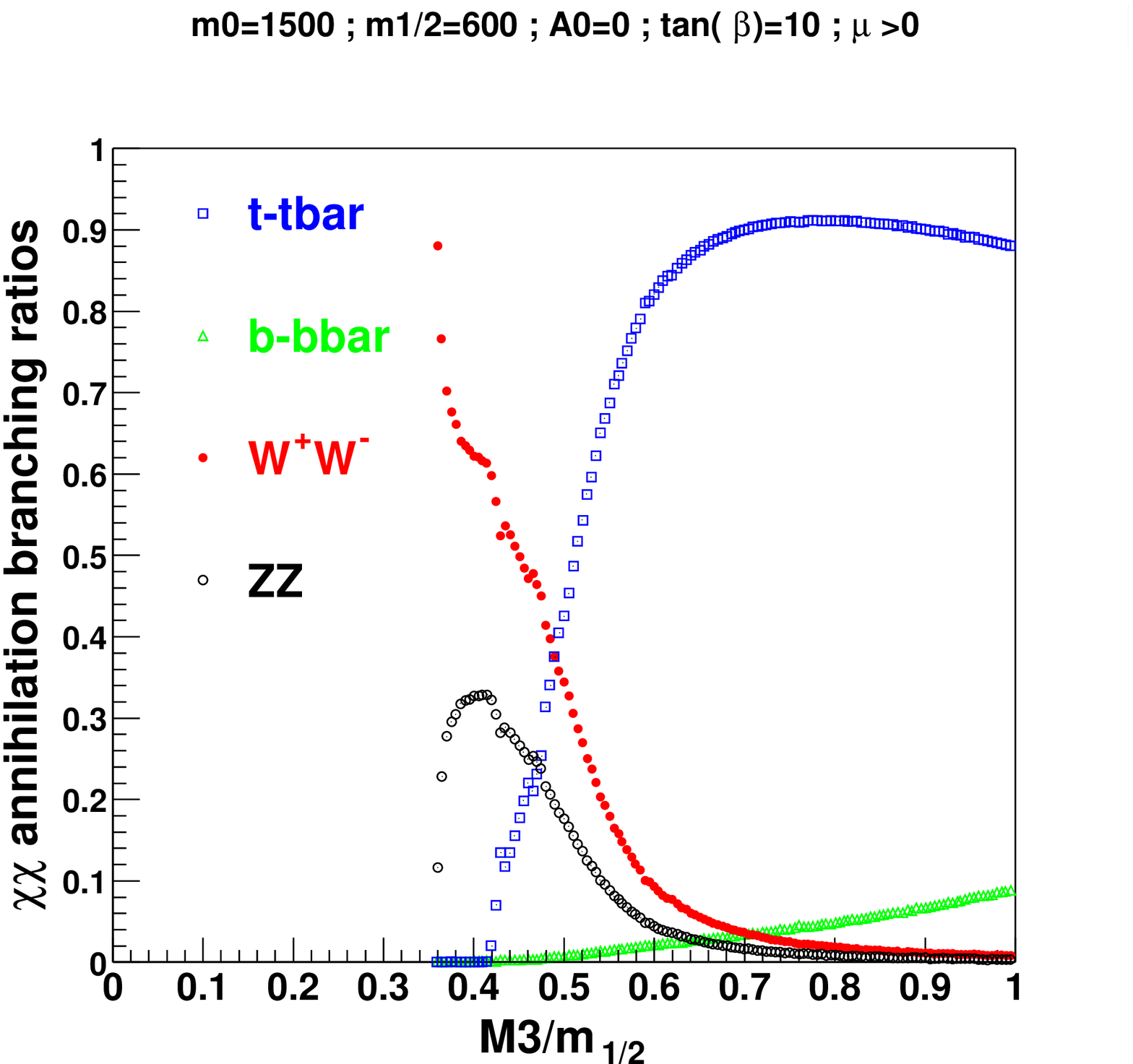}&
\includegraphics[width=0.33\textwidth]{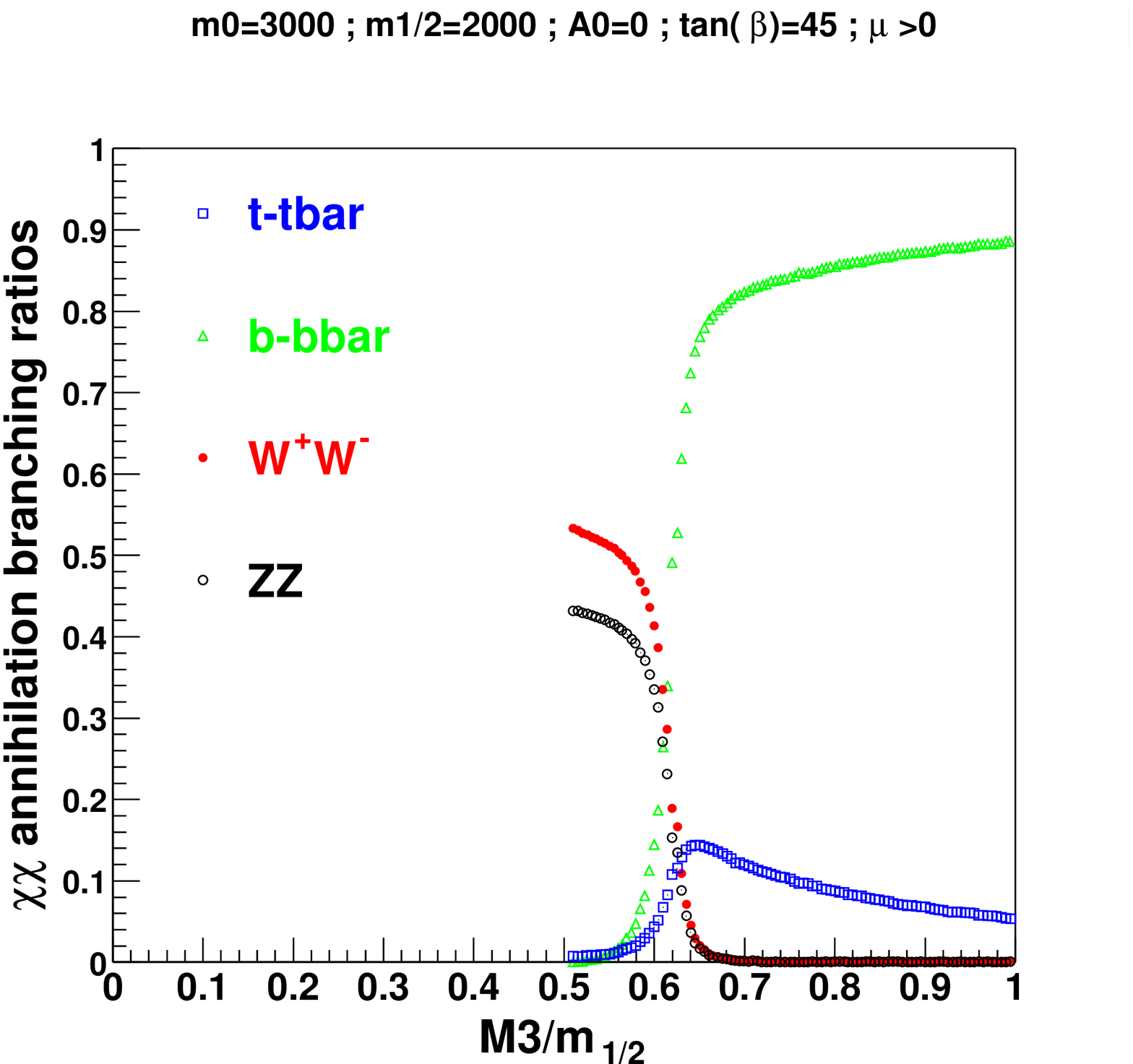}\\
a) & b) & c)
\end{tabular}
\caption{\small Evolution of the dominant neutralino annihilation
  branching ratios as functions of the $M_3/m_{1/2}$ ratio for the models
  a) of figures~\ref{M3M2effect}, b) of figures~\ref{M3M2effect3} and c) of
  figures~\ref{M3M2effect2}.}
\label{M3bratios}
\end{center}
\end{figure}

For the CMSS Model with $m_0=1500$ {\rm GeV}, $m_{1/2}=600$ {\rm
GeV}, $A_0=0$ {\rm GeV}, $\tan{\beta}=45$, $\mu>0$ (figures
\ref{M3M2effect} and \ref{M3bratios}a ), the dominant channel at
$x=1$ is $\chi\chi\xrightarrow{A}b\bar{b}$. By decreasing $x$, the
latter remains at first dominant while the neutralino relic density
is reduced due to the decrease of $m_A$ (figure \ref{M3M2effect}a and
f ), then since $\mu$ also decreases, the processes
$\chi\chi\xrightarrow{Z}t\bar{t}$ followed by
$\chi\chi\xrightarrow{\chi^+_i} W^+W^-$ and
$\chi\chi\xrightarrow{\chi_i} ZZ$ take successively over (as well as
the $\chi\chi^+$ and $\chi\chi^0_2$ coannihilations) further lowering
the relic density. The enhancement of the $t\bar{t}$ branching ratio
before the increase of the neutralino higgsino fraction ($x>0.8$) (see
figure \ref{M3bratios}a ) is due to the decrease of the stop mass.

For the same CMSS Model but with $\tan{\beta}=10$ (figures
\ref{M3M2effect3} and \ref{M3bratios}b ), $m_A$ is larger due to
the smaller $\tan{\beta}$ value, so the
$\chi\chi\xrightarrow{Z}t\bar{t}$ channel dominates the neutralino
annihilation. When $x$ is lowered, this process remains dominant
(figure \ref{M3bratios}b ) but its cross section firstly increases
because $m_{\tilde{t}}$ decreases. Then the $Z$ exchange process
proportional to the neutralino higgsino fraction takes over due to the
decrease of $\mu$ (see figure \ref{M3M2effect3}d ) enhancing the
annihilation cross section and decreasing the relic density (see
figure \ref{M3M2effect3}a ). Finally, the annihilation processes into
gauge bosons become dominant for $x\lesssim 0.5$ because $m_{\chi}$ becomes
smaller than the top mass.

For the CMSS Model with $m_0=3000$ {\rm GeV}, $m_{1/2}=2000$ {\rm GeV},
$A_0=0$ {\rm GeV}, $\tan{\beta}=45$, $\mu>0$ (figures \ref{M3M2effect2} and
\ref{M3bratios}c ), the neutralino relic density is much too large in the
CMSSM, but lowering $M_3$ drives this model into the cosmologically favoured
region by the decrease of $m_A$ and $\mu$ (figures \ref{M3M2effect2}a,
\ref{M3M2effect2}d and \ref{M3M2effect2}f ). This model is located further
away from the radiative electroweak symmetry breaking boundary and has
larger $\mu$. In addition, $m_A<m_{\tilde{q}}$ so the process
$\chi\chi\xrightarrow{A}b\bar{b}$ strongly dominates over
$\chi\chi\xrightarrow{Z,\tilde{t}}t\bar{t}$. Again, annihilations into
gauge bosons take over when $x$ ({\it i.e.} $\mu$) decreases.

\noindent
{\bf Neutralino direct and indirect detections:}\\
When $x$ is lowered, the direct detection yields (see figures
\ref{M3M2effect}c, \ref{M3M2effect3}c and \ref{M3M2effect2}c ) can be
enhanced by 2 or 3 orders of magnitude with respect to the CMSSM case
($x=1$). Firstly the reduction of the squark masses favours the $\chi q
\xrightarrow{\tilde{q}} \chi q$ process, and mainly the coupling $C_{\chi q
  H}$ is maximal for maximum $z_{13(4)}z_{11(2)}$ mixed products which
increase with the neutralino higgsino fraction when $x$ decreases.
Moreover, $m_H\sim m_A$ and thus $\sigma^{scal}_{\chi-p}$ is enhanced due
to the decrease of $m_A$. When the neutralino gaugino fraction finally
drops, the $z_{13(4)}z_{11(2)}$ products decrease and
$\sigma^{scal}_{\chi-p}$ decreases back with $x$. This behaviour can
clearly be remarked on the figures \ref{M3M2effect}c, \ref{M3M2effect}e,
\ref{M3M2effect}f, \ref{M3M2effect3}c, \ref{M3M2effect3}e,
\ref{M3M2effect3}f and \ref{M3M2effect2}c, \ref{M3M2effect2}e,
\ref{M3M2effect2}f.

As far as neutrino indirect detection is concerned, the enhancement on the
muon fluxes coming from neutralino annihilation in the Sun, due to the
decrease of $M_3|_{GUT}$, can reach up to 6 orders of magnitude with
respect to the CMSSM case ($x=1$). The main effect is coming from the
increase of the spin dependent neutralino-proton elastic cross section,
firstly due to the decrease of the squark masses which enhance $\chi
q\xrightarrow{\tilde{q}}\chi q$ in $\sigma^{spin}_{\chi-p}$, then mainly to
the decrease of $\mu$ leading to a larger higgsino fraction in the
neutralino which enhance $\chi q\xrightarrow{Z}\chi q$ in
$\sigma^{spin}_{\chi-p}$. Moreover the larger higgsino fraction also favours
the neutralino annihilations into the $\chi\chi\rightarrow W^+W^-,\ ZZ$ and
$\chi\chi\rightarrow t\bar{t}$ channels which give harder neutrino spectra
than $\chi\chi\rightarrow b\bar{b}$. This is illustrated on figures
\ref{M3M2effect}b, \ref{M3M2effect}d, \ref{M3M2effect}e,
\ref{M3M2effect3}b, \ref{M3M2effect3}d, \ref{M3M2effect3}e and
\ref{M3M2effect2}b, \ref{M3M2effect2}d, \ref{M3M2effect2}e. This
enhancement is not as peaked in $x$ as for direct detection but remains
maximum as long as the higgsino fraction dominates the neutralino
composition. However the relic density becomes very small when $x$ is
further lowered.  For these extreme low values of $M_3$ with small relic
density, the enhancement of the muon fluxes coming from neutralino
annihilation in the centre of the Earth (enhanced by
$\sigma^{scal}_{\chi-p}$ and $\sigma^A_{\chi-\chi}$) can become significant
but usually remains beyond reach of current and next generation neutrino
telescopes' sensitivities.

To summarise, we have shown it is possible, by lowering $M_3|_{GUT}$, to
decrease the neutralino relic density to the desired cosmological value for
any CMSS Model even for $m_0$ and $m_{1/2}$ as large as several TeV. The
value of $M_3|_{GUT}$ necessary to get a relic density $\Omega_{\chi} h^2
\sim0.1-0.2$ mainly depends on $m_{1/2}$ : $0.5 m_{1/2}<M_3<m_{1/2}$.

In the CMSSM, the main neutralino annihilation channels can be regrouped in
two sets: the pseudo-scalar exchange $\chi\chi\xrightarrow{A}b\bar{b}$ and
the processes directly proportional to the neutralino higgsino fraction
$\chi\chi\xrightarrow{Z}t\bar{t}$, $\chi\chi\xrightarrow{\chi^+_i} W^+W^-$
and $\chi\chi\xrightarrow{\chi_i} ZZ$. As we have seen, the decrease of the
$M_3|_{GUT}$ parameter influences many MSSM parameters through the RGE and
leads to {\it both } a decrease of $m_A$ {\it and} an increase of the
neutralino higgsino fraction by the decrease of $\mu$. So the dominant
neutralino annihilation process of any CMSS Model can be enhanced in order
to obtain the good value for the relic density.

\begin{figure}[t!]
\begin{center}
\begin{tabular}{cc}
\multicolumn{2}{c}{{ $A_0=0\ ;\ \tan{\beta}=45\ ;\ \mu>0\ ; M_3/m_{1/2}=0.63$}}\\
\multicolumn{2}{c}{\includegraphics[width=\textwidth]{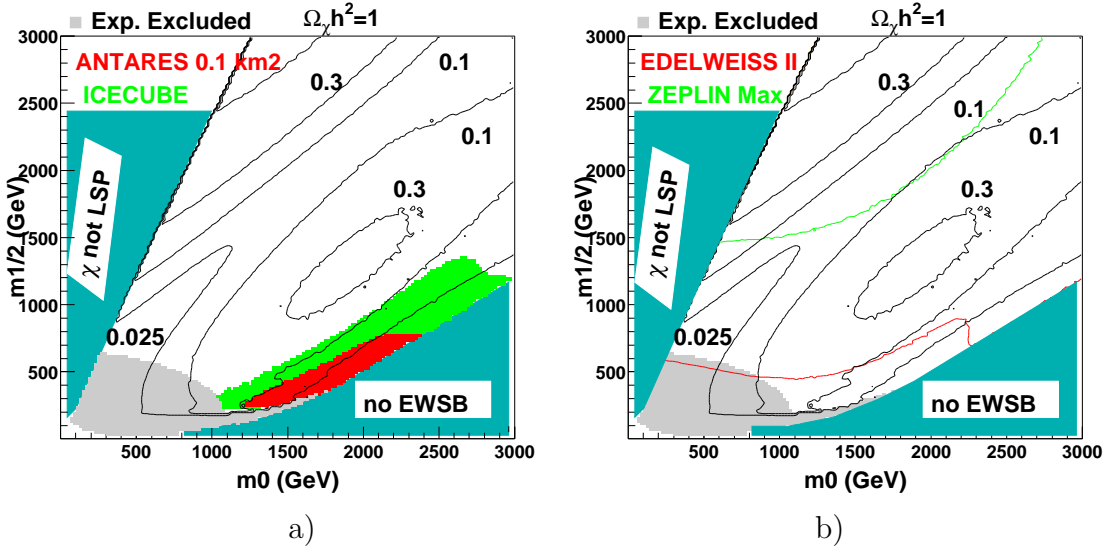}}\\
\hspace{0.25\textwidth}a) & \hspace{0.15\textwidth}b)
\end{tabular}
\caption{\small Neutralino detection potential in the $(m_0,m_{1/2})$
plane for $x=M_3/m_{1/2}=0.63$ for a) neutrino telescopes looking at
muon fluxes coming from the Sun and b)
direct detection experiments.}
\label{x063}
\end{center}
\end{figure}

Varying the $x$ value allows to have $\Omega_{\chi} h^2\sim0.15$ in the
whole shown $(m_0,m_{1/2})$ plane. For $\tan{\beta}<50$, where the $A$ pole
is not present in the neutralino annihilation, the interesting values of
$x$ range between $\sim 0.8-1$ along the zone with a correct neutralino
relic density in the CMSSM and down to $x\sim0.5-0.6$ for large $m_{1/2}>2000$
GeV and $m_0<2000$ GeV. At the centre of the shown ($m_0$,$m_{1/2}$) plane, a
quite generic $x$ value is found
\begin{equation}
M_3|_{GUT}\sim 0.6(\pm0.1)m_{1/2}+{\rm corrections}(m_0,\tan{\beta},m_b)
\end{equation}
permitting to get close to the $A$ pole or at least to adjust the couple
$(m_A,z_{11(2)}z_{13(4)})$ and thus the $\chi\chi\xrightarrow{A}b\bar{b}$
channel in order to get $\Omega_{\chi} h^2 \sim0.1-0.2$. These $x$ values
also favour the direct detection yields through the $\chi
q\xrightarrow{H}\chi q$ process. Moreover, the decrease of $M_3$ increases
the neutralino higgsino fraction, so that a typical value
$M_3/m_{1/2}\lesssim0.8$ enlarges the ``focus-point'' corridor at large
$m_0$ as well as the region of the parameter space accessible to direct and
indirect detection. Notice that all $x$ values discussed above lie well
above $x=0.16$, below which the gluino becomes the LSP.

\begin{figure}[t!]
\begin{center}
\begin{tabular}{cc}
\multicolumn{2}{c}{{ $A_0=0\ ;\ \tan{\beta}=10\ ;\ \mu>0\ ; M_3/m_{1/2}=0.55$}}\\
\multicolumn{2}{c}{\includegraphics[width=\textwidth]{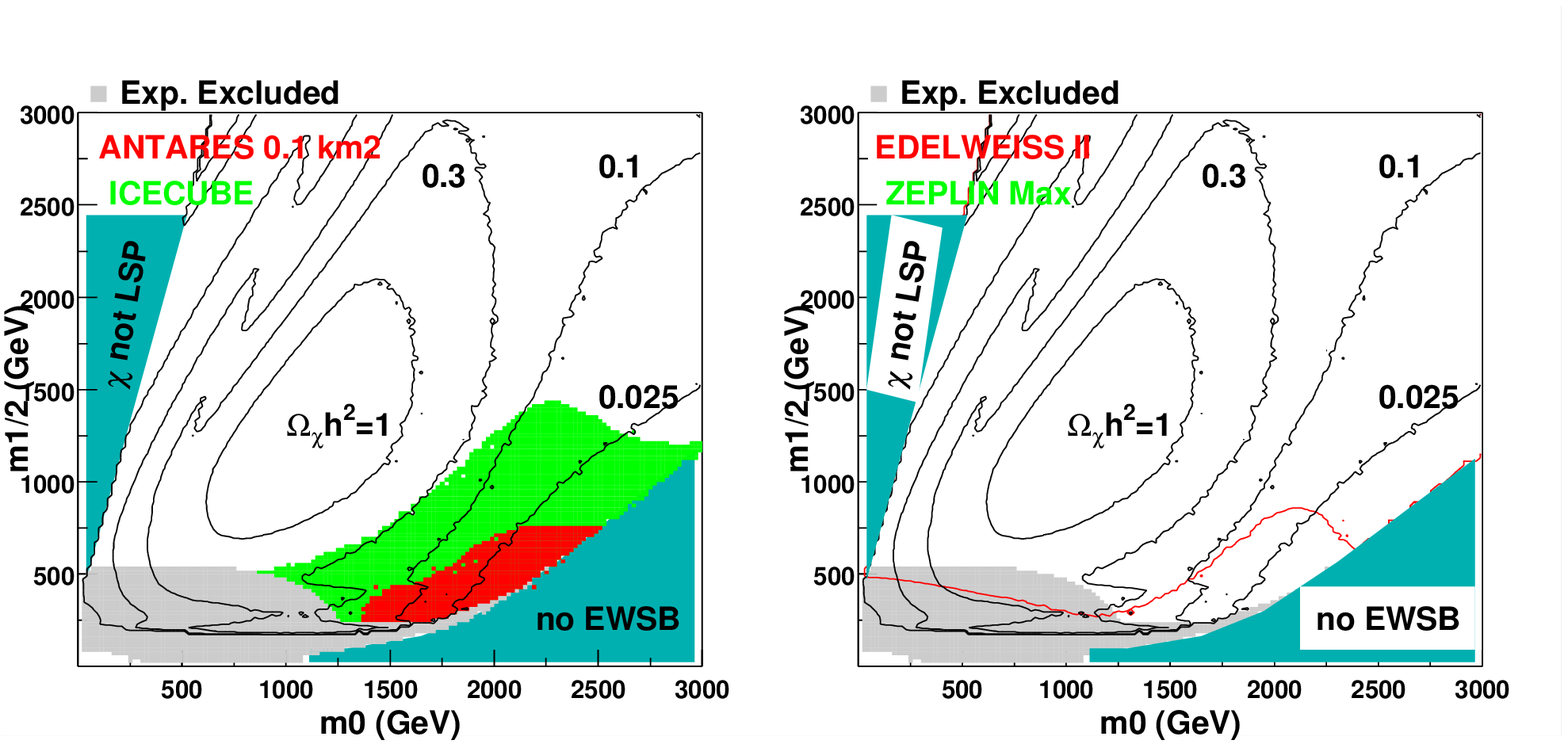}}\\
\hspace{0.25\textwidth}a) & \hspace{0.15\textwidth}b)
\end{tabular}
\caption{\small Neutralino detection potential in the $(m_0,m_{1/2})$
  plane for $\tan{\beta}=10$ with $x=M_3/m_{1/2}=0.55$ for a) neutrino
  telescopes looking at muon fluxes coming from the Sun and b) direct
  detection experiments (For this value of $x$, Zeplin Max covers the whole
  shown plane).}
\label{x055tanb10}
\end{center}
\end{figure}

Two examples with a fixed ratio $M_3/m_{1/2}$ are shown on figures
\ref{x063} and \ref{x055tanb10} where regions with interesting relic
density and experiment sensitivity areas are vastly improved with respect
to the CMSSM case \cite{Bertin:2002ky}, especially noticing the larger
$m_{1/2}$ range shown.

\noindent
{\bf Links to SUSY breaking :}\\
\noindent
From the above discussions, $M_3|_{GUT}$ appears as the most relevant
parameter in SUSY non-universal models as far as neutralino detection is
concerned and we have defined its departure from universality by the
$x=M_3|_{GUT}/m_{1/2}$ ratio. In line with the logics of minimising the
deviations to universality, we can map this single parameter $x$ into a
single combination of the {\bf 24}, {\bf 75} and {\bf 200} representations:
\begin{equation}
{\rm \hbox{non-univ}\ =\ CMSSM}\ + c_{24}{\bf 24}+c_{75}{\bf
75}+c_{200}{\bf 200} 
\end{equation}
which leads for the relative values of the gaugino masses
($M_1:M_2:M_3$) at $M_{GUT}$ to
\begin{equation}
(1:1:x)_{{\rm non\ univ}} = (1:1:1)_{{\rm
CMSSM}}+(0:0:x-1)_{c_{24}{\bf 24}+c_{75}{\bf 75}+c_{200}{\bf 200}}.
\end{equation}

From table \ref{SU5relations}, we then see that the coefficients must take
the values:
\begin{equation}
c_{24}=\frac{20}{63}(x-1)\ ;\ c_{75}=\frac{14}{63}(x-1)\ ;\
c_{200}=\frac{9}{63}(x-1).  
\end{equation}
giving for the typical value of interest $x=0.6$ : $c_{24}\simeq-0.13$ ;
$c_{75}\simeq-0.09$ ; $c_{200}\simeq-0.06$, corresponding to quite small
coefficients indeed.

A possible origin of such lower $M_3$ values might otherwise be found in
anomaly mediated SUSY breaking (AMSB) schemes.  Even if the usual
derivation of AMSB spectra leads to wino LSP's
\cite{Moroi:1999zb,Huitu:2002fg}, other patterns have been found
\cite{Chen:1997ap} where the gluino can become lighter, to the point of
having a gluino LSP in extreme cases. An exploration of this connection in
the general framework of \cite{Binetruy:2000md} would be interesting.

\section{Low energy effective parameterisation of the MSSM}

We have explored the grand unification SUSY models favourable to the
indirect detection of neutralino dark matter with neutrino telescopes. This
kind of study has previously been performed in the low energy MSSM
\cite{Jungman:1996df,Edsjo:1997hp}. It is obvious that such a low energy
approach offers more free parameters (in particular $\mu$) to play with,
and it thus becomes much easier to obtain heavy neutralino with a strong
higgsino component. The main difference between the grand unification and
the low energy approaches comes from the absence of neutralino annihilation
into Higgs bosons for the GUT models, due to the RGE evolution of the SUSY
parameters and to the mass spectrum hierarchy obtained. This possibility of
lighter scalars (Higgs and squarks of first generation) in the low energy
approach also sometimes enables fluxes from the Earth to be significant.
However these low energy effective models are less consistent and do not
take into account some nice features of SUSY GUT models such as the
radiative electroweak symmetry breaking or the avoidance of Landau poles
and CCB minima.

\begin{figure}[t!]
\begin{center}
\psfrag{musurM1}[c][r]{$\mu/M_1$}
\psfrag{M1}{$M_1\ [{\rm GeV}]$}
\psfrag{0}[c][c]{0}
\psfrag{200}[c][c]{200}
\psfrag{400}[c][c]{400}
\psfrag{600}[c][c]{600}
\psfrag{800}[c][c]{800}
\psfrag{1000}[c][c]{1000}
\psfrag{1}[r][r]{1}
\psfrag{1.2}[r][r]{1.2}
\psfrag{1.4}[r][r]{1.4}
\psfrag{1.6}[r][r]{1.6}
\psfrag{1.8}[r][r]{1.8}
\psfrag{2}[r][r]{2}
\psfrag{higgsino}[l][l]{higgsino}
\psfrag{fraction}[l][l]{fraction}
\psfrag{=}[l][l]{=}
\psfrag{,}[l][l]{,}
\psfrag{0.1}[l][l]{0.1}
\psfrag{0.2}[l][l]{0.2}
\psfrag{0.3}[l][l]{0.3}
\psfrag{0.4}[l][l]{0.4}

\includegraphics[width=0.6\textwidth]{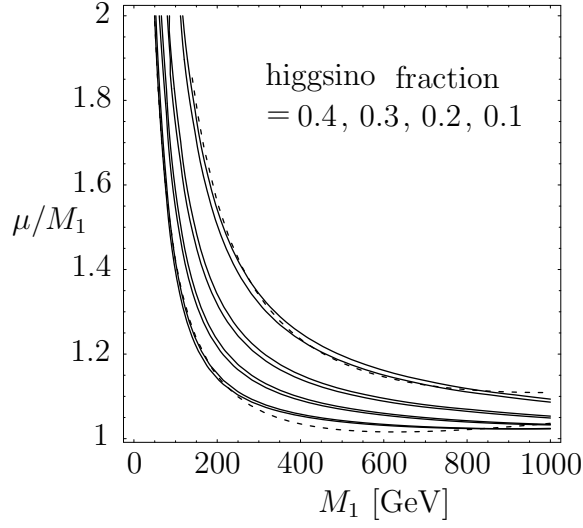}
\caption{Analytical iso higgsino fractions  for $\tan{\beta}=10$ 
  (lower twin curve) and 50 (upper twin curve) shown as full lines, and the
  fitted functions described in equation (\ref{mufit}) for $f_H=0.4$ (lower
  left) and $f_H=0.1$ (upper right) shown as dashed lines.}
\label{hfrac}
\end{center}
\end{figure}

We propose here to derive a low energy effective parameterisation of
SUSY GUT models which appear to be favourable for the neutrino
indirect detection of neutralinos. All these models present indeed the
same characteristics:
\begin{itemize}
\item a non-vanishing neutralino higgsino fraction $f_H$ :  $0.1\lesssim f_H
\lesssim 0.4$ (see figures \ref{resumemSugra} and \ref{M3M2effect},
\ref{M3M2effect3}, \ref{M3M2effect2}),
\item $\chi\chi\xrightarrow{\chi^+_i,\chi_i} W^+W^-,\ ZZ$ or
$\chi\chi\xrightarrow{Z}t\bar{t}$ dominant neutralino annihilation
channels which are very efficient in order to get both a good value of the
neutralino relic density and hard neutrino spectra.
\end{itemize}
One notices that all the physics which governs these processes only
involves neutralinos and/or charginos on top of standard
particles. Therefore a minimal description assumes all scalars to be heavy and
decoupled. The remaining necessary parameters are those defining the
neutralino sector : $M_1$, $M_2$, $\mu$ and $\tan{\beta}$. Moreover the
unification relations and the RGE evolutions lead to $M_2\simeq 2 M_1$,
while a neutralino higgsino fraction in the range $0.1\lesssim f_H \lesssim
0.4$ implies (see figure \ref{hfrac})

\begin{equation}
1.4\times10^{-4}M_1^2+0.83M_1+60 \lesssim \mu 
\lesssim 1.5\times10^{-4}M_1^2+0.8M_1+150
\label{mufit}
\end{equation}
with negligible dependence in $\tan{\beta}$ (see twin curves of figure \ref{hfrac}).\\

\begin{figure}[ht!]
\begin{center}
  \psfrag{hf01}{$f_{H}\sim 0.1$} \psfrag{hf04}{$f_{H}\sim 0.4$}
  \includegraphics[width=0.8\textwidth]{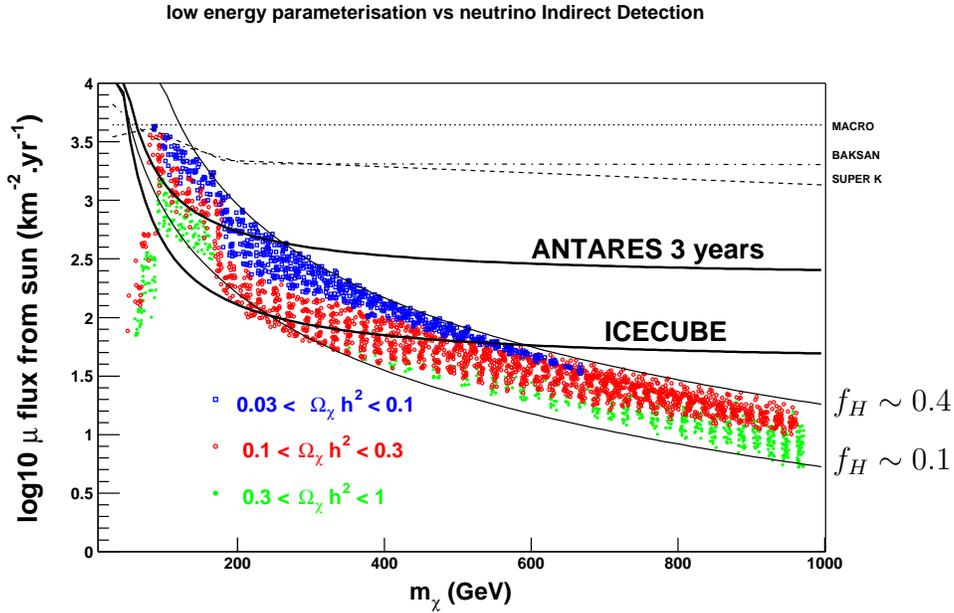}
\caption{Low energy parameterisation of RGE models versus experiment 
  sensitivities. The thin full lines are the fitted functions of equation
  (\ref{mufluxfit}).}
\label{paramscan}
\end{center}
\end{figure}

A SUSY GUT model satisfying at low energy: $M_{{\rm scal}}\gg M_1$,
$M_2\simeq 2M_1$ and $\mu$ in the range indicated by equation (\ref{mufit})
gets a cosmologically favoured value of the neutralino relic density and
large muon fluxes coming from neutralino annihilation in the Sun. For a
scan of low energy models with $50<M_1<1000$ GeV, $M_{{scal}}\gg M_1$,
$\tan{\beta}=5,10,20,30,40,50,60$, and $\mu$ constrained by equation
(\ref{mufit}) to get $0.1\lesssim f_H \lesssim 0.4$, we find the following
range for the solar $\mu$ flux (see figure \ref{paramscan})
\begin{equation}
f_H\sim0.1 \leftrightarrow 5\times 10^6 
  \frac{e^{50/m_{\chi}}}{m_{\chi}^2}
\lesssim log_{10}[\mu\ {\rm flux}_{\odot}]
\lesssim 1.5 \times 10^7 \frac{e^{180/m_{\chi}}}{m_{\chi}^2}
\leftrightarrow f_H\sim0.4.
\label{mufluxfit}
\end{equation}

We thus see that independently of their high energy scale realisation, the
most favourable models are within reach of the next generation (${\rm
  km^3}$) of neutrino telescopes through the detection of neutrino fluxes
from the Sun, up to masses of the order of 450 GeV for the most interesting
relic densities.
 
\section{Conclusion}

In this paper, we have explored departures from CMSSM universality and
singled out those most interesting for neutralino dark matter detection.
The scalar sector can be of interest to adjust the neutralino relic density
by opening sfermions coannihilation processes, but does not lead to
increased detection rates because of the RGE evolution of first generation
soft terms. The most determining parameters are the gaugino masses
$M_2|_{GUT}$ and particularly $M_3|_{GUT}$ which respectively increase the
wino and the higgsino content of the neutralino when lowered away from
their universal values. The higgsino component is more efficient than the
wino one to improve the detection rates, making of $M_3|_{GUT}$ the most
relevant degree of freedom, as its value also affects the whole MSSM
spectrum \cite{Kazakov:1999pe}. Such models with lower $M_3|_{GUT}$ values
have a better relic abundance and are much more promising from a detection
point of view, with rates increased by several orders of magnitude compared
to the universal case. Some naturalness objections could be addressed on
the high values of $m_0$ and $m_{1/2}$ commonly used in the literature.
However it should be noticed that lower values of $M_3$ tend to be more
natural \cite{Kane:1998im}. To finish, all the models with large neutrinos
fluxes studied here tend to have small SUSY contribution to the $\mu$
anomalous magnetic moment ($a^{\rm SUSY}_{\mu} \simeq 0$), which might
exclude them, but not before the $a_{\mu}$ value of the Standard Model
itself is clearly ruled out.

\acknowledgments{We thank the "GdR Supersym\'etrie" of the French CNRS for
  its support, and in particular acknowledge the early fruitful exchanges
  with L.
  Duflot\footnote{http://duflot.home.cern.ch/duflot/GDR/GPS\_audela/GPS\_audela.html}
  inside of the working group "Au del\`a de l'universalit\'e" initiated by
  J.-F. Grivaz, whose influential role shows up all the way to the title of
  the present work. More recently, we would also like to thank G. Belanger,
  S. Rosier-Lees and R. Arnowitt for useful discussions.}

\nocite{}
\bibliography{nonuniv}
\bibliographystyle{JHEP}
\end{document}